\documentstyle[12pt,epsf]{article}
\makeatletter
\def\@cite#1#2{{[{#1}]\if@tempswa\typeout
{IJCGA warning: optional citation argument
ignored: `#2'} \fi}}
\newcount\@tempcntc
\def\@citex[#1]#2{\if@filesw\immediate\write\@auxout{\string\citation{#2}}\fi
  \@tempcnta\z@\@tempcntb\m@ne\def\@citea{}\@cite{\@for\@citeb:=#2\do
    {\@ifundefined
       {b@\@citeb}{\@citeo\@tempcntb\m@ne\@citea\def\@citea{,}{\bf ?}\@warning
       {Citation `\@citeb' on page \thepage \space undefined}}%
    {\setbox\z@\hbox{\global\@tempcntc0\csname b@\@citeb\endcsname\relax}%
     \ifnum\@tempcntc=\z@ \@citeo\@tempcntb\m@ne
       \@citea\def\@citea{,}\hbox{\csname b@\@citeb\endcsname}%
     \else
      \advance\@tempcntb\@ne
      \ifnum\@tempcntb=\@tempcntc
      \else\advance\@tempcntb\m@ne\@citeo
      \@tempcnta\@tempcntc\@tempcntb\@tempcntc\fi\fi}}\@citeo}{#1}}
\def\@citeo{\ifnum\@tempcnta>\@tempcntb\else\@citea\def\@citea{,}%
  \ifnum\@tempcnta=\@tempcntb\the\@tempcnta\else
   {\advance\@tempcnta\@ne\ifnum\@tempcnta=\@tempcntb \else
\def\@citea{--}\fi
    \advance\@tempcnta\m@ne\the\@tempcnta\@citea\the\@tempcntb}\fi\fi}
\makeatother

\newcommand{\gsim}{\lower.7ex\hbox{$\;\stackrel{\textstyle>}{\sim}\;$}}
\newcommand{\lsim}{\lower.7ex\hbox{$\;\stackrel{\textstyle<}{\sim}\;$}}
\newcommand{\be}{\begin{equation}}
\newcommand{\ee}{\end{equation}}
\newcommand{\bea}{\begin{eqnarray}}
\newcommand{\eea}{\end{eqnarray}}

\def\baselinestretch{1}
\begin{document}
%%%%%%%%%%%%%%%%%%%%%%%%%%% subequations.sty %%%%%%%%%%%%%%%%%%%%%%%%
\catcode`@=11
\newtoks\@stequation
\def\subequations{\refstepcounter{equation}%
\edef\@savedequation{\the\c@equation}%
  \@stequation=\expandafter{\theequation}%   %only want \theequation
  \edef\@savedtheequation{\the\@stequation}% % expanded once
  \edef\oldtheequation{\theequation}%
  \setcounter{equation}{0}%
  \def\theequation{\oldtheequation\alph{equation}}}
\def\endsubequations{\setcounter{equation}{\@savedequation}%
  \@stequation=\expandafter{\@savedtheequation}%
  \edef\theequation{\the\@stequation}\global\@ignoretrue

\noindent}
\catcode`@=12
%%%%%%%%%%%%%%%%%%%%%%%%%%%%%%%%%%%%%%%%%%%%%%%%%%%%%%%%%%%%%%%%%%%%%
\begin{titlepage}
\vspace{5mm}
\title{{\bf Preheating curvature perturbations with a coupled curvaton}}  
\vskip2in
\author{  
{\bf M. Bastero-Gil$^{1}$\footnote{\baselineskip=16pt E-mail: {\tt
mbg20@pact.cpes.susx.ac.uk}}},
{\bf V. Di Clemente$^{2,3}$\footnote{\baselineskip=16pt E-mail: {\tt
vicente@thphys.ox.ac.uk}}}
and
{\bf S. F. King$^{3}$\footnote{\baselineskip=16pt E-mail: {\tt
sfk@hep.phys.soton.ac.uk}}}  
\hspace{3cm}\\   
$^{1}$~{\small Department of Physics and Astronomy, University of
Sussex,} \\  
{\small Falmer, Brighton, BN1 9QH, U.K.}
\hspace{0.3cm}\\
$^{2}$~{\small Theoretical Physics, University of Oxford, 1 Keble Road,} \\ 
{\small Oxford, OX1 3NP, U.K.}
\hspace{0.3cm}\\
$^{3}$~{\small Department of Physics and Astronomy, University of Southampton,} \\ 
{\small Highfield, Southampton, SO17 1BJ, U.K.}
\hspace{0.3cm}\\
}
\date{} 
\maketitle 
\def\baselinestretch{1.15} 
\begin{abstract}
\noindent 
We discuss the potentially important r\^{o}le played
by preheating in certain variants of the curvaton mechanism
in which isocurvature perturbations of a D-flat (and F-flat) direction
become converted to curvature perturbations during reheating.
We analyse the
transition from inflation to reheating in some detail, including the
dynamics of the coupled curvaton and inflation fields during this
transition. We discover that preheating could be an important source
of adiabaticity  
where parametric resonance of the isocurvature components amplifies
the super-horizon fluctuations by a significant amount.
As an example of these effects we develop 
a particle physics motivated model which we recently introduced in
which the D-flat direction is identified with the usual Higgs field. Our new
results show that it is possible to achieve the correct curvature
perturbations for initial values of the curvaton fields of order the
weak scale. 
In this model we show that the prediction for the spectral index of the final
curvature perturbation only depends on 
the mass of the curvaton during inflation,
where consistency with current
observational data requires the ratio of this mass to the
Hubble constant to be $\leq 0.3$. 
\end{abstract}

\thispagestyle{empty}
\vspace{2cm}
\leftline{\today}
\leftline{}

%This needs fixing now
\vskip-23.5cm
\rightline{}
\rightline{SHEP 03-30}
%\rightline{hep-ph/0311237}

\end{titlepage}

%%%%%%%%%%%%%%%%%%%%%%%%%%%%%%%%%%%%%%%%%%%%%%%%%%%%%%%%%%%%%%%%%%%

\setcounter{footnote}{0} \setcounter{page}{1}
\newpage
\baselineskip=20pt

\noindent

\section{Introduction}

In a recent paper we proposed a new mechanism
for generating curvature perturbations during reheating in hybrid
inflation \cite{Bastero-Gil:2002xr}. The idea was that, during inflation,
one of the slowly rolling scalar fields, other than the inflaton,
could develop large isocurvature perturbations, and that this
could become converted to curvature perturbations during the process
of reheating. This is quite different to the curvaton mechanism
as originally proposed \cite{lyth1,moroi} since the scalar with large
isocurvature perturbations is not late decaying\footnote{For recent
discussion on several scenarios for the late decaying    
curvaton see Refs. \cite{curvaton1,curvaton2,curvaton3,curvaton4}).}. 
Instead the mechanism 
in \cite{Bastero-Gil:2002xr} relies on the simple observation that 
just after the end of inflation, at the onset of reheating,
the fields of hybrid inflation \cite{hybrid,copeland} tend to 
share out the vacuum energy which dominates during inflation.
This typically leads to hybrid fields with energy densities 
of the same order of magnitude during the epoch of reheating, 
allowing efficient conversion of isocurvature to curvature
perturbations during reheating. This mechanism should not be
confused with the late decaying curvaton mechanism, whose
physics is not related to reheating. However for ease of terminology
we shall continue to call the hybrid field whose isocurvature
perturbations during inflation become converted to curvature 
perturbations during reheating the ``curvaton'', even though
our mechanism is different from the curvaton mechanism as 
originally proposed.

The r\^{o}le of our curvaton will be played by a D-flat (and F-flat)
direction\footnote{In supersymmetric models, a flat direction has to
be both D-flat and F-flat. For the inflaton and the $N$ mediating
field, being singlets, the D-flatness condition is trivial fulfilled. 
By calling the curvaton a D-flat direction what we want to stress is
its non-singlet nature.} during
inflation, {\rm coupled} to the waterfall field
$N$ of hybrid inflation. Therefore it is a ``coupled curvaton'',
which again is different from the original curvaton idea.
Furthermore, being a D-flat direction, typically it will decay through
its gauge interaction before the singlets whose decay rate is
controlled by small Yukawa couplings. We gave an explicit
example of the general mechanism in which the
the Higgs fields of the supersymmetric standard model
played the r\^{o}le of a {\it coupled curvaton}, i.e. developed 
isocurvature perturbations during inflation which
subsequently become converted to curvature perturbations
\cite{Bastero-Gil:2002xr}. 
We discussed the nature and evolution of the Higgs perturbations
during the epoch of inflation, and described the evolution of the
perturbations during reheating after the Higgs decayed. 
After Higgs decay, the Universe consists of a mixture of matter (the 
oscillating singlet fields) and radiation (Higgs decay products),
and we were able to show that the isocurvature perturbations are converted
into curvature perturbations in this framework.
However in the previous paper
we did not analyse the transition from inflation to reheating
before the Higgs curvaton decayed, and we also neglected the
very important effects of preheating.

The purpose of the present paper is to discuss both of the above
effects in detail, with surprising results (at least to us).
Due to the presence of the {\it coupled curvaton}, 
the isocurvature perturbation (entropy)
between the inflaton and the $N$ waterfall field is converted into the
adiabatic curvature {\em during the transition from inflation to reheating}.
The value of the
total curvature perturbation at the end of the transition from 
inflation to reheating is then given
in terms of the value of the coupled curvaton field rather than the
inflaton. As it turns out, 
with the coupled curvaton in hybrid inflation, the
initial value at horizon crossing of the isocurvature perturbation
{\it decreases} during the transition from inflation to reheating. 
To be precise,
we find that during the transition from inflation to reheating,
before the Higgs curvaton decays, the curvature perturbation drops
by several orders of magnitude from our estimate in 
\cite{Bastero-Gil:2002xr}. Fortunately when preheating is taken
into account the curvature perturbation may subsequently be increased 
by a similar number of orders of magnitude, resulting in
an acceptable final amplitude of curvature perturbations,
consistent with COBE and WMAP. 

The layout of the remainder of the paper is as follows.
In section 2 we briefly review the model,
and in section 3 we describe the inflationary trajectory and the
transition from inflation to reheating. In section 4 we describe 
the evolution of the fluctuations during inflation, 
and we carefully analyse the effect of the transition from 
inflation to reheating. It is during this transition that the
curvature perturbation first changes. In section 5 we discuss the
effects of preheating on the  curvature perturbation, 
while allowing the Higgs to decay to radiation. Section 6 contains a
discussion of the prediction of the spectral index and its correlation
with the parameters which are consistent with giving correct
structure. Section 7 concludes the paper.

\section{The inflationary model}

%\subsection{The model} 
The supersymmetric hybrid inflation
model is based on the superpotential \cite{steve}:
\be
W= \lambda N H_u H_d - \kappa \phi N^2 \,, 
\label{superpot}
\ee
where $N$ and $\phi$ are singlet superfields, $H_{u,d}$ are the Higgs
superfields, and $\lambda$, $\kappa$ are dimensionless couplings.
Other cubic terms in the superpotential are forbidding by imposing a
global $U(1)_{PQ}$ Peccei-Quinn symmetry. 
The superpotential in Eq. (\ref{superpot}) includes a linear
superpotential for the inflaton field, $\phi$, typical of hybrid inflation,
as well as the singlet $N$ coupling to Higgs doublets as in the NMSSM.
We notice that in standard hybrid inflation \cite{hybrid,copeland} the
singlet $N$ will quickly settle to its false vacuum value at zero,
leaving the inflaton flat direction to slow-roll until it reaches the
critical value. This means that during inflation the $\mu$ term
generated by the coupling of $N$ to the Higgses in (\ref{superpot})
vanishes, and the Higgses become an F-flat direction. Therefore, we
can have an additional flat direction made by a 
combination of the Higgs fields  which satisfies the D-flatness
condition \cite{kolda}, i.e., such that D-term contribution to the
potential $(g_2^2+g_1^2) ( H_u^2-H_d^2)^2/8$ vanishes; for example
taking them to be equal, $H_u=H_d=h$. Therefore, a more general
inflationary trajectory is given by allowing the Higgs fields to take
non-vanishing, although as we will see, small values.

Inflation takes place below the SUSY breaking
scale. Including the soft SUSY breaking masses, $m_\phi$,
$m_N$ and $m_h$, and trilinears $A_\kappa$, $A_\lambda$,  
the potential for the real part of the fields reads\footnote{The
axionic (imaginary) part of the complex fields can be set consistently
to zero, and it will not affect the evolution of the real scalar
fields either during inflation or the reheating period.}: 
\bea
V &=& V(0) +  \frac{\kappa^2}{4} N^4 + \frac{\lambda^2}{4} h^4 +
\kappa^2(\phi-\phi_c^+)(\phi-\phi_c^-)N^2 
 + \frac{\lambda^2}{2}N^2 h^2   +
\lambda N h^2 \left(\frac{A_\lambda}{\sqrt{2}} - \kappa \phi \right)
\nonumber \\
&{}& + \frac{1}{2} m_\phi^2 \phi^2 + \frac{1}{2} m_h^2 h^2 +
\frac{1}{2} m_N^2 N^2  \,,  
\label{V}
\eea
The constant $V(0)$ has been added to ensure the necessary vanishing of
the potential at its global minimum, and the critical value of $\phi$,
$\phi_c^\pm$, are given below in terms of the soft SUSY parameters:
\be
\phi_c^\pm =
\frac{A_\kappa}{2\sqrt{2}\kappa}
\left(1\pm\sqrt{1-\frac{4m_N^2}{A_\kappa^2}}\right)   
\,. 
\label{phic}
\ee 
As we will see in the next section, during inflation (when
$\phi>\phi_c^+$) the inflaton field, $\phi$, is slowly rolling along
the potential and  
also the fields $N$ and $h$ get small values but such that the $N$
field dependent mass squared, $\bar{m}_N^2$, is dominated mainly by the term
$\kappa^2(\phi-\phi_c^+)(\phi-\phi_c^-)$. Once at the critical value,
$\phi = \phi_c^+$, $\bar{m}_N^2$ changes sign and both singlets, $\phi$ and
$N$, roll down towards the global minimum $\phi_0$ and $N_0$ ending
inflation\footnote{We 
shall see that $h\ll N_0, \phi_0$ and also that $m_\phi^2, m_N^2\ll
\kappa^2 N_0^2$. Therefore either contributions from the Higgs field, $h$, and
from the bare masses, $m_\phi^2, m_N^2$, can be neglected in order to
calculate the global minima, $\phi_0$ and $N_0$.}.
\bea
\phi_0 &\approx& \frac{\phi_c^+ + \phi_c^-}{2} =
\frac{A_\kappa}{2\sqrt{2}\kappa} \, , \label{phi0} \\
N_0 &\approx& \frac{\phi_c^+ - \phi_c^-}{\sqrt{2}} =
\frac{A_\kappa}{2\kappa}\sqrt{1-\frac{4 m_N^2}{A_k^2}}\, . \label{N0}
\eea
The Higgs field, $h$, will oscillate once inflation ends around the minima
which is zero as long as the Electroweak Symmetry is still
preserving\footnote{In a future work~\cite{future} we will show how
the electroweak symmetry breaking is implemented in this model and
their cosmological consequences. However here
we are going to consider that the electroweak symmetry is
preserved all the time, since our principal conclusions will not be
affected by this issue.}. 

Before describing the inflationary trajectory, 
we briefly discuss the values of the parameters of the potential
in Eq. (\ref{V}). They are uniquely determined by embedding this model into
an extra-dimensional framework with just one fundamental (``string'') scale
$M_*\sim 10^{13}$ GeV \cite{Bastero-Gil:2002xs}. Embeding
all the Higgs fields and singlets $\phi$ and $N$ in the bulk, while all the
matter fields live on the brane, it is possible to show
that~\cite{Bastero-Gil:2002xs,future} 
\bea
&{}& \lambda \sim \kappa \sim \left(\frac{M_*}{m_P}\right)^2 \sim
10^{-10}\, ,
\label{coupling} \\
&{}& A_\kappa \sim A_\lambda \sim M_* \left(\frac{M_*}{m_P}\right)^2 \sim
10^{3} \,\mbox{GeV}\,, \label{trilinear} \\
&{}& m_\phi \sim m_N \sim m_h \sim \frac{1}{4\pi} M_*
\left(\frac{M_*}{m_P}\right)^3 \sim 1 \,\mbox{MeV}\,,
\label{inflaton_mass} \\ 
&{}& V(0)^{1/4} \sim M_*\left(\frac{M_*}{m_P}\right) \sim 10^8
\,\mbox{GeV}\,, \label{V0} 
\eea
where $m_P=M_{P}/\sqrt{8 \pi}=2.4\times10^{18} \,\mbox{GeV}$ is the reduced
Planck mass. Using Eqs. (\ref{phi0}) and (\ref{N0}) we obtain:
\be
\phi_0 \sim N_0 \sim M_* \sim 10^{13} \mbox{GeV} \, ,
\label{minima}
\ee
and the Hubble parameter during inflation is of the order of
\be
H = \frac{V(0)^{1/2}}{\sqrt{3} m_P} \simeq 10\, MeV \,. 
\ee
Notice that $m_\alpha < H$ for $\alpha=\phi,N,h$.

We will use those values parameters in the following sections when
presenting  
numerical  results. However the mechanism
is general for a potential like Eq. (\ref{V}) independently of
parameter values, and  the analytical estimations 
are presented in general in terms of the model scales\footnote{For order of
magnitude estimations, we will take $\phi_c \sim \phi_c^\pm \sim
\phi_0 \sim N_0$, and $\lambda \simeq \kappa$.}, $H$, $\phi_c$,
$\kappa \phi_c$. We will refer in the following to the ``Higgs'' like
the coupled curvaton, D-flat direction during inflation, included in
Eq. (\ref{V}).      

\section{Evolution of the background fields.}

Including the Higgs terms in the potential Eq. (\ref{V}), 
the term proportional to $\lambda N h^2$
can induce a non-zero value for $h$ and $N$ during inflation providing that 
the coefficient of this term is negative
$\frac{A_\lambda}{\sqrt{2}}-\kappa \phi<0$. 
However, in order to ensure a flat direction for the Higgs, and
therefore allowing it to slow-roll during inflation,
its effective field dependent mass must be smaller than $H^2$. The
latter receives contributions from all the fields, 
\bea
\frac{\partial^2 V}{\partial h^2} & \simeq & \lambda^2 (3 h^2 + N^2) + 2
\lambda N( \frac{A_\lambda}{\sqrt{2}}- \kappa \phi) + m_h^2 < H^2 \,.
%\frac{\partial^2 V}{\partial \phi^2} & \simeq & \kappa^2 N^2 + m_\phi^2 <
%H^2 
\eea
This implies that both the fields $N$ and $h$ must
take small values during inflation, with 
\be 
\frac{N}{\phi_c} < \left(\frac{H}{\kappa \phi_c}\right)^2 \approx
10^{-10} \,,\;\;\;\;\; \frac{h}{\phi_c} < \frac{H}{\lambda \phi_c}
\approx 10^{-5}\,, 
\label{small}
\ee
so that $N \ll h \ll \phi$. These values are small enough not to
disturb the inflaton slow-roll, 
\bea
\frac{\partial^2 V}{\partial \phi^2} & \simeq & \kappa^2 N^2 +
m_\phi^2 \simeq m_\phi^2 < H^2 \,. 
\eea
The Higgs and inflaton fields evolve independently of each other and
$N$, following the approximate slow-roll trajectories given by:
\be
\dot \phi(t) \simeq - \eta_\phi H \phi(t) \,, \label{dotphi} \;\;\;\;\;\;
\dot h(t) \simeq - \eta_h H h(t) \,,
\ee
where $\eta_h= m_h^2/(3 H^2)$,  $\eta_{\phi}= m_{\phi}^2/(3 H^2)$.
Inflation as such is still controlled by the inflaton $\phi$, 
and will end when  $\phi$ reaches one of its critical values $\phi_c^{\pm}$. 

On the other hand, the $N$ field follows the evolution equation:
\be
\ddot N + 3 H \dot N + V_N = 0 \,,
\label{evolN}
\ee
with 
\bea
V_N=\frac{\partial V}{\partial N}&=& ( \kappa^2 N^2+ 2 \kappa^2
(\phi-\phi_c^+)(\phi-\phi_c^-) + \lambda^2 h^2) N + \lambda h^2
(\frac{A_\lambda}{\sqrt{2}}- \kappa \phi ) \nonumber \\
&\approx & \omega_N^2(\phi) N +
\lambda h^2 (\frac{A_\lambda}{\sqrt{2}}- \kappa \phi )  \,, 
\label{dVN}
\eea   
where in the second line $\omega_N^2(\phi)\gg H^2$ is the
field dependent effective squared $N$ mass, where we have dropped the
small term $\kappa^2N^2+ \lambda^2 h^2$ using
Eq. (\ref{small}). During inflation we can take the 
slow-rolling fields $h$ and $\phi$ to be approximately constant in
Eq. (\ref{dVN}). Thus, the $N$ field will oscillate with an
amplitude damped by the expansion as $\propto a(t)^{-3/2}$ and
a frequency of the order of $\omega_N(\phi) \simeq O(\kappa \phi_c)
\gg H$, but around a non vanishing value given by the quasi-constant
contribution in Eq. (\ref{dVN}), i.e,
\be
N(t) \simeq \frac{ \lambda h^2}{\omega_N^2(\phi)}
(\kappa \phi -\frac{A_\lambda}{\sqrt{2}})  + \frac{1}{a(t)^{3/2}}
F_{osc}[\omega_N(\phi)]  \,, 
\label{Nosc}
\ee
where $F_{osc}[\omega_N(\phi)]$ represents the oscillating part. 
Given that in any case the amplitude of the oscillations quickly 
decays due to the exponential expansion, after some e-folds they
become completely negligible, and Eq. (\ref{Nosc}) gives the trajectory:
\be
N(t) \simeq  \frac{ \lambda h^2}{\omega_N^2(\phi)}
( \kappa \phi- \frac{A_\lambda}{\sqrt{2}}) \,,
\label{Nh}
\ee
that is, the motion of $N$ effectively follows the valley of minima in
the potential given  by $V_N=0$.
This inflationary trajectory for $N$ is independent of
the initial conditions, provided that the other fields are already 
slowly rolling. 
Although in origen $N$ is a ``heavy'' field (mass larger than the Hubble
rate), the dynamics is such that it behaves effectively as a ``light''
one. The quasi-exponential expansion  during inflation settles it to
the minimum, but the minimum is a  
time-dependent, slow-rolling one. This has implications also from the
point of view of its quantum fluctuations: they will behave in a
similar way as those of a light field. 

%\subsection{Around the critical value and towards the minimum} 
\begin{figure} 
\begin{tabular}{cc} 
\epsfxsize=8cm
\epsfysize=8cm
\epsfbox{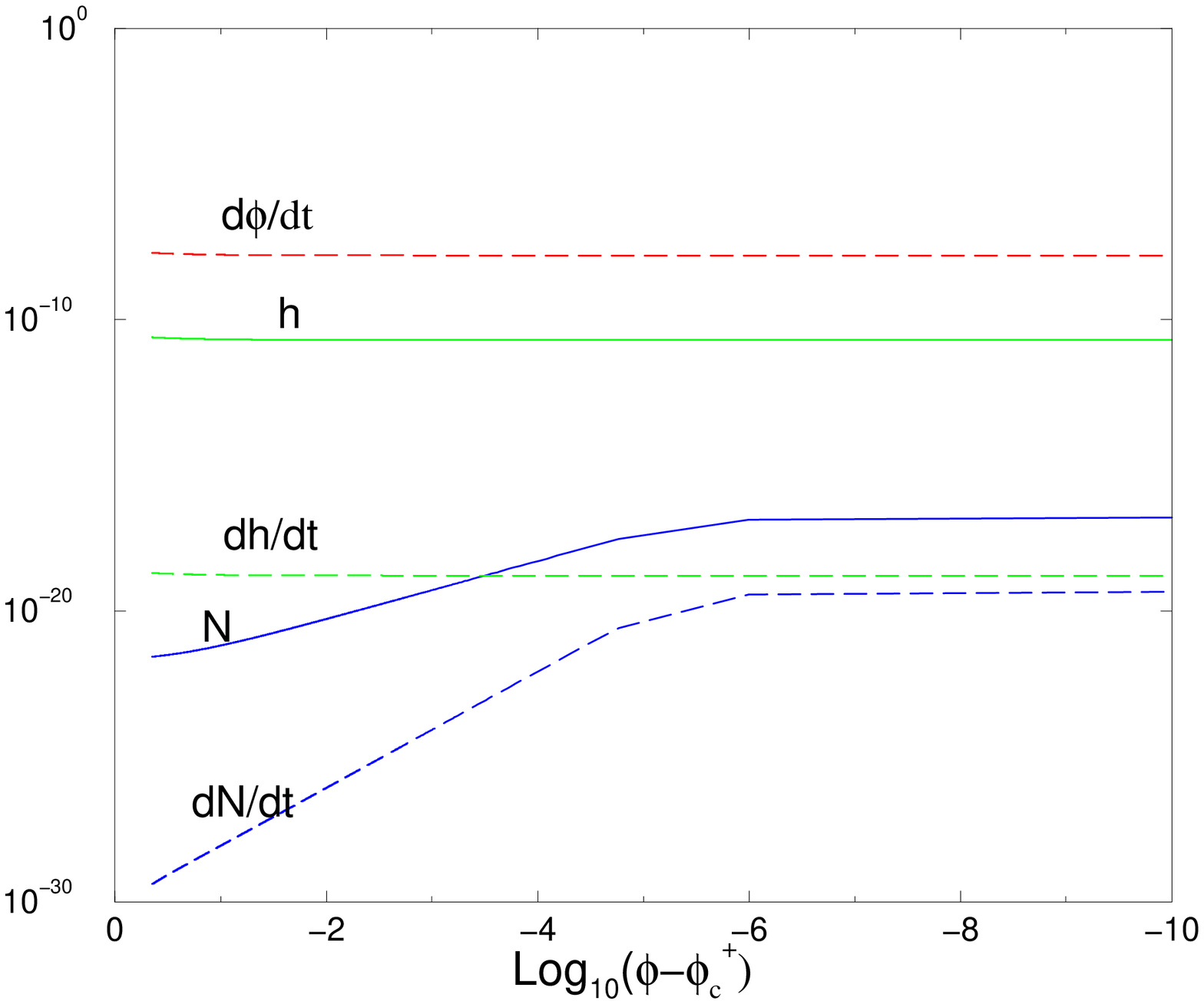}
&
\epsfxsize=8cm
\epsfysize=8cm
\epsfbox{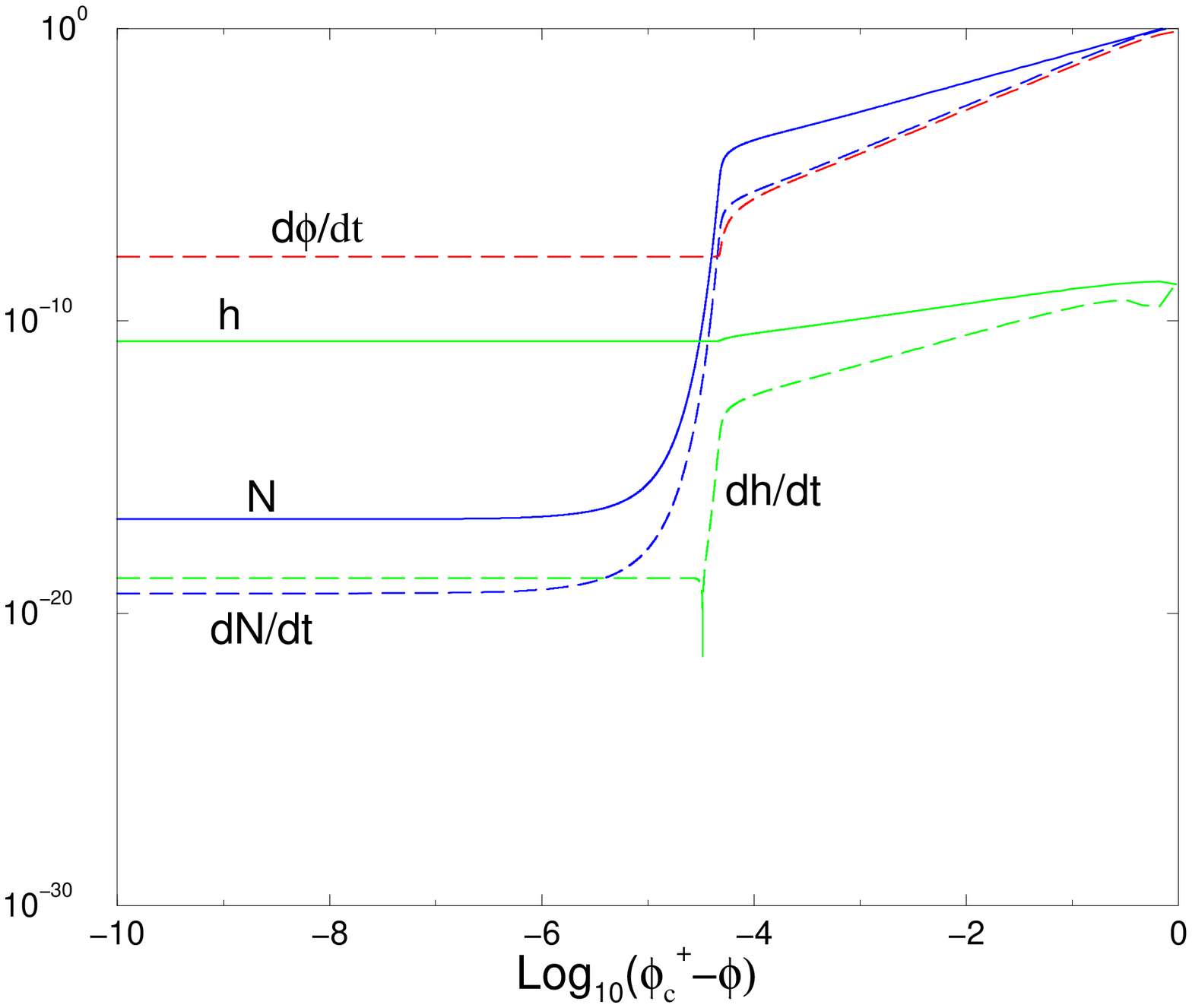} 
\end{tabular} 
\caption{\footnotesize We show the evolution of the background fields
(solid lines) 
and time derivaties (dashed lines) during inflation up to the critical
value $\phi=\phi_c^+$ (LHS),  and from the critical value up to
the global minimum (RHS). We have taken the values of the parameters:
$\phi_0=10^{13}$ GeV, $\phi_c^+=2 \phi_0$, $\kappa \phi_0=1$ TeV,
$A_\lambda=1.5 \sqrt{2} \kappa \phi_0$, $m_\phi=m_h=0.1 H$, and the
value of the Higgs field $h(0)= 247$ GeV. The
numerical integration starts around 60 e-folds before the end of
inflation. \label{fig0}}
\end{figure} 

The evolution of the background fields and their time derivaties
can be seen in Fig. (\ref{fig0}). We have choosen to plot
the evolution 
against the diference $|\phi - \phi_c^+|$ in order to detail the
changes around the critical point, which we analyse in the
remaining of this section. In the left panel, we have plotted the
evolution during inflation up to the critical point, and in the right
one that from the critical point to the global minimum. Time flows in
both from left to right, and we have started the numerical integration
around 60 e-folds before the end of inflation. Although the $N$ field is also
``evolving'' during inflation, this is standard hybrid inflation in
the sense that it is the $N$ field which triggers its end  as can be
seen in RHS plot in Fig. (\ref{fig0}), when $\phi$ leaves the
slow-rolling trajectory $\dot \phi \approx$ Constant. 

In the LHS plot in Fig. (\ref{fig0}), it
can be seen that close to but before $\phi$ reaches the critical
value, as  $\omega_N^2(\phi)\rightarrow 0$, the 
values of the derivatives of $N$ increase very rapidly and the
effective ``slow roll'' of the $N$ field given in Eq. (\ref{Nh}) is
violated. In
particular, $\ddot N$  becomes now the dominant term to
cancel out the 
Higgs and $\phi$ quasi-constant contribution in the equation of
motion, such that 
\be
\ddot N + 3 H \dot N + V_N=0 \Rightarrow \ddot N \simeq \lambda h^2
(\kappa \phi-\frac{A_\lambda}{\sqrt{2}} ) \label{ddotN}
\,.    
\ee
We want to estimate the ratio $(\dot N/N)$ at the critical value
$\phi=\phi_c^+$, which will be of use later when computing the metric
perturbations.   
The transition around the critical value, and from there to the global
minimum, takes a very short interval of time with $H \Delta t \ll 1$,
so that we can safely neglect the effect of the expansion in the
analytical estimations. By matching the above  trajectory
Eq. (\ref{ddotN}) with the 
inflationary one Eq. (\ref{Nh}), we 
see that the transition between trajectories occurs when 
\be
\kappa(\phi-\phi_c^+)_- \simeq \left( \frac{\dot  
\phi}{\phi}\right)^{2/3}\left( \frac {(\kappa \phi_c^+)^2}{\kappa
(\phi_c^+-\phi_c^-)} \right)^{1/3} \,,   
\label{deltaphic}
\ee
and we have denoted this point by a subindex ``$-$''. Given the
short interval of time lapsed between this point and the critical 
value, practically 
the values of $N$ and $\dot N$ at
$\phi_c^+$ are given by those derived from Eqs. (\ref{Nh}) but evaluated
at $\kappa (\phi-\phi_c^+)_-$ above, which gives the ratio, 
\be
\left( \frac{\dot N}{N} \right)_c \simeq
\frac{\eta_\phi H \phi_c^+  }{(\phi-\phi_c^+)_-}
\simeq \left( \eta_\phi H
\kappa^2(\phi_c^+-\phi_c^-) \phi_c^+ \right)^{1/3} \,, 
\label{dotNNc}
\ee
where we have used  the slow-roll approximation\footnote{This is only
consistent if by that time 
the value of $N_c$ is still small enough  not to disturb the
others trajectories, i.e,
\be
N_c \simeq \frac{\lambda h^2 (\kappa \phi_c^+-
A_\lambda/\sqrt{2})} 
{ (\kappa \phi_c^+ \kappa (\phi_c^+-\phi_c^-))^{2/3}} \left(\frac{
\phi}{\dot \phi}\right)^{2/3} < \left(\frac{H}{\kappa
\phi_c^+}\right)^2 \phi_c^+ 
\ee
which implies
\be
\lambda h < O(H \left(\frac{\dot \phi}{\phi}{\kappa \phi_c}
\right)^{1/3}) < H \,,
\label{condh}
\ee  
which we assume to hold hereafter unless otherwise stated. 
That is, end of inflation in the standard sense will proceed only
after $\phi$ crosses  the critical value. If Eq. (\ref{condh}) were
not fulfilled, for example when the value of the Higgs at horizon
crossing is very close to the upper limit given in Eq. (\ref{small}),
then inflation will terminate instead  before $\phi$ reaches
the critical value by the increasing values of $N$ and $h$.} 
 $\dot \phi/\phi \simeq \eta_\phi H$. 

Then, as shown in the RHS plot in Fig. (\ref{fig0}), as soon as
$\phi$ crosses the critical value and the $N$ field dependent 
squared mass becomes negative, the value of the field starts growing
exponentially, 
\be
N \sim N_c exp( \int \sqrt{2 \kappa (\phi_c^+-\phi) (\phi-\phi_c^-)}
dt)\,, 
\label{Nexp}
\ee
until finally the large values of $N$ modify the slow-roll of
the inflaton field, and this also moves fast towards the minimum. 
Afterwards, $N$ and $\phi$ will move towards a straight-line 
trajectory \cite{us} given by
\be
N = \sqrt{2} (\phi_c^+ - \phi) \label{straight}\,. 
\ee  
Once $N$ reaches the trajectory in Eq. (\ref{straight}), the slow-roll
regime for the inflaton (and the Higgs) ends.  
The Higgs also gets desestabilized due to the increasing $N$ field, and 
 it will just become increasingly coupled to the others, but at a
slower pace;  
i. e., the Higgs does not really affect the transition of the inflaton
and the $N$ from the critical point to the global minimum.

\section{Curvature perturbation during inflation and the subsequent 
phase transition.}  

Our final aim is to compute the spectrum of primordial curvature
perturbations, initially originated during inflation.
In the presence of several light (slow-rolling) scalar fields
$\phi_\alpha$, the comoving curvature perturbation \cite{lukash} can
be written as \cite{gordon,gordon2}    
\be
{\cal R} = 
H \sum_\alpha \left( \frac{ \dot \phi_\alpha}{\sum_\beta
\dot \phi_\beta^2} \right) Q_\alpha\,,  
\label{RQ} 
\ee
in terms of the gauge-invariant scalar field amplitude
perturbations, the Sasaki-Mukhanov variables $Q_i$ \cite{Qi},
\be
Q_\alpha = \delta \phi_\alpha + \frac{\dot \phi_\alpha}{H} \psi, 
\label{Qi}
\ee
where $\delta \phi_\alpha$ is the quantum fluctuaction of the
background field $\phi_\alpha$, and $\psi$ the gauge-dependent metric
perturbation (see Appendix A). 
The total curvature perturbation is then given as the
weighted sum of ``individual'' curvature perturbations 
\be
{\cal R}_\alpha= H \frac{Q_\alpha}{\dot \phi_\alpha} \,,
\ee
such that,
\be
{\cal R}= \sum_\alpha \frac{\dot \phi_\alpha^2}{\sum_\beta
\dot \phi_\beta^2} {\cal R}_\alpha = \sum_\alpha \frac{\rho_\alpha
+P_\alpha}{\rho +P} {\cal R}_\alpha\,, 
\label{curdef}
\ee 
with $\dot \phi_\alpha^2= \rho_\alpha + P_\alpha$ for a scalar
field, $\rho_\alpha$ being the energy density and $P_\alpha$ the
pressure. The RHS of Eq. (\ref{curdef}) applies in general to a
multi-component Universe, independently of the nature (scalar fields
or other) of the components. In order to follow the evolution of the
curvature  perturbation from inflation to the reheating stage, we
will follow in detail that of the individual components ${\cal
R}_\alpha$ and how their relative contribution to ${\cal R}$ changes. 

During inflation, nevertheless, the field that dominates the total
curvature perturbation is the one with the largest velocity, this
being the inflaton field $\dot \phi \gg \dot h \gg \dot N$. From
Eq. (\ref{curdef}), we simply have  ${\cal R}_{inf} \simeq {\cal R}_{\phi}$. 
However, as we will see next, the curvature perturbations of the Higgs
and $N$, ${\cal R}_h$ and ${\cal R}_N$, are orders of magnitude larger
than that of  ${\cal R}_\phi$. Relative perturbations between
different components 
are the origen of the so called entropy or isocurvature\footnote{So
called as far as the originating field or fluid component does not
dominate the total energy density, and therefore its fluctuations make
little contribution to the curvature one.} perturbations,
${\cal R}_\alpha - {\cal R}_\beta$. Therefore
in our scenario, given the strong hierarchy between the values of the
fields,  there is a large ``isocurvature'' perturbation between
the inflaton and the Higgs, and the inflaton and the $N$. Those can later
seed the total curvature perturbation, if one of
the field (kinetic) energy 
density other than that of the inflaton becomes non-negligible 
\cite{silvia,mukhanov,polarksi,deruelle,juandavid,riotto}.  
This is what happens at the end of inflation: as soon as the inflaton
field reaches the critical value, the field $N$ evolution changes and
moves {\it fast} towards the global minimum, with increasing kinetic
energy.  Therefore, during the
transition from 
the critical point $\phi_c^+$ towards the minimum $\phi_0$, the field
$N$ with a large kinetic energy gives the dominant contribution to the total
curvature perturbation in Eq. (\ref{curdef}); i.e., the entropy
perturbation between $N$ and $\phi$ becomes adiabatic curvature
perturbation. This can be seen by writting the evolution
equation for the total curvature\footnote{ As we are only interested
in large scale fluctuations, we neglect the term that goes like $(k/aH
\ll 1)$ in the equation.} in the presence of the scalar fields
\cite{juandavid} $\phi$, $N$ and $h$:
\bea
\dot{\cal R} &\simeq& -3 H \sum_{\alpha,\beta}(c_\alpha^2 -
c_\beta^2)\frac{h_\alpha  h_\beta}{h_T^2}({\cal R}_\alpha- {\cal
R}_\beta) \nonumber \\ 
&\simeq& -3 H (c^2_\phi - c^2_N) \frac{h_\phi h_N}{h_T^2}
({\cal R}_\phi -   {\cal R}_N) - 3 H (c^2_N - c^2_h) \frac{h_N h_h}{h_T^2}
({\cal R}_N -   {\cal R}_h) \nonumber \\
&-& 3 H (c^2_h - c^2_\phi) \frac{h_h h_\phi}{h_T^2} 
({\cal R}_h -   {\cal R}_\phi) \,, 
\label{dotRscalars}
\eea
where as a shorthand notation we have defined: $h_\alpha=
\rho_\alpha+p_\alpha=\dot \phi_\alpha$; and $h_T$ is the total density
energy, $h_T = h_\phi + h_N + h_h$; $c_\alpha^2=\dot
p_\alpha/\dot \rho_\alpha$ is the 
sound speed for each component,
\be
c_\alpha^2= 1 + \frac{2 V_\alpha}{3 H \dot \alpha}\,. 
\label{calpha}
\ee
Clearly when $\dot N \simeq \dot \phi$, the first term
in Eq. (\ref{dotRscalars}) will dominate, such that ${\cal R} \simeq
{\cal R}_N$. Notice also that during the transition $\phi$ and $N$
behave as ``different fluid'' components with {\it
different sound speeds}: the former is still slowly-rolling while $N$
is increasing exponentially.  

For a multi-scalar system, rotating the fluctuations in field space,
as done in Ref. \cite{gordon2}, helps to identify all along the
``adiabatic'' mode  responsible for the the total curvature, and the
``entropy'' modes  
that influence its evolution. For comparison with this approach, we
have included the rotated equations in Appendix B. 
Entropy modes source the adiabatic one
when the fields follow a curved trajectory in field space. This
matches the analitical behaviour in the next subsection: during the
transition, the trajectory is curved by the motion of $N$, such that
the curvature changes.

\subsection{During inflation.} 
The equation of motion for the gauge invariant quantum fluctuation
$Q_\alpha$, with comoving wavenumber $k$, is given by \cite{taruya}  
\be
\ddot Q_\alpha + 3 H \dot Q_\alpha + \frac{k^2}{a^2}Q_\alpha + \sum_\beta
V_{\alpha \beta} Q_\beta= 
\left[ \frac{1}{a^3 m^2_P} \left( \frac{a^3}{H} \dot \phi_\alpha \dot
\phi_\beta \right)^{\cdot } \right] Q_\beta \,,
\label{ddotQalpha}
\ee
where $V_{\alpha_\beta}= \partial^2 V/\partial \phi_\alpha
\partial \phi_\beta$. For the inflaton and Higgs fields 
we have $V_{\phi\phi}$, $V_{hh}$, $V_{\phi h} \ll H^2$ during
inflation; i.e., their 
fluctuacions behave nearly as those of a massless field. Hence,
$Q_\phi$ and  $Q_h$ will be frozen to a constant value 
once outside the horizon, $k < aH$, given
approximately by the value at horizon crossing $k=aH$ \cite{inflation}, 
\be
Q_{\phi *}= Q_{h *}=\frac{H_*}{\sqrt{2 k^3}} \label{qi}\,,
\ee
with ${\cal R}_\phi= H Q_\phi/\dot \phi \ll {\cal R}_h= H
Q_h/\dot h$ during inflation. 

On the other hand, neglecting for simplicity the
sub-dominant\footnote{The metric contributions are of the order of
$O(\dot \phi_\alpha^2/\rho)$, and therefore negligible during
inflation.}  terms coming
from metric contributions on the R.H.S in Eq. (\ref{ddotQalpha}), the
evolution equation for $Q_N$ reads     
\be
\ddot Q_N + 3 H \dot Q_N + (\frac{k^2}{a^2} + V_{NN} ) Q_N  
 + V_{N\phi} Q_\phi + V_{Nh} Q_h \simeq 0 \,.
\ee 
Like in the case of  the evolution of the background field $N$, 
now the large mass term $V_{NN} \sim O( \kappa^2 \phi_c^2) \gg H^2$ gives 
rise to oscillations with an amplitude decaying as $a^{-1}$, but
displaced from zero due to the mixing with $Q_\phi$ and $Q_h$. 
Therefore, once the amplitude of the oscillations
becomes negligible, the $N$ field
fluctuation will also tend to a quasi-constant value given by
\be
Q_N \simeq -\frac{V_{N\phi}}{V_{NN}} Q_\phi - \frac{V_{Nh}}{V_{NN}} Q_h \,,
\ee
and using Eq.~(\ref{Nh}) we have for $Q_N$ and the background velocity
$\dot N$:
\bea
Q_N &\simeq& N \left( 2 \frac{Q_h}{h} + C_\phi(\phi)
\frac{Q_\phi}{\phi} \right)\simeq 2 N \frac{Q_h}{h} \,, \label{QN}\\
\dot N &\simeq& N \left( 2 \frac{\dot h}{h} + C_\phi(\phi) \frac{\dot
\phi}{\phi} \right) \label{dotN}\,, 
\eea
with 
\be
C_\phi(\phi) = \frac{\kappa \phi}{\kappa \phi - A_\lambda/\sqrt{2}} - \frac{4
\kappa^2(\phi-\phi_0)\phi}{\omega_N^2(\phi)} \sim O(1)\,, 
\ee
and $Q_h/h \gg Q_\phi/\phi$. 
Then, using Eqs. (\ref{QN}) and (\ref{dotN}), ${\cal R}_N$ is given
during inflation by: 
\bea 
{\cal R}_N &\simeq& 2 \left(\frac{N}{\dot N} \right) \left(\frac{\dot
h}{h} \right) {\cal R}_h \label{RNinf} \\
 &\simeq& -\frac{H}{2 \eta_h + C_\phi(\phi) \eta_\phi}
\frac{Q_h}{h}\,,
\eea
which is of the same order of magnitude than ${\cal R}_h$ during inflation, 
while $\phi$ is not too close to the critical value. We stress
that it is the Higgs flat direction which holds the $N$ field to a
non-vanishing, non-decreasing value during
inflation, in a kind of ``slow-roll'' trajectory; in a similar way 
their quantum fluctuations $Q_N$ follow $Q_h$ instead of being
redshifted away. 

We do not require any special initial 
conditions for Eq. (\ref{RNinf}) to hold, apart from the fields being
already in their slow-roll trajectories. Given that inflation will
last  more than this, the 
fields will have most probably reached that trajectory much before the
largest observable scale today become super-Hubble, around say 60
e-folds before the end of inflation. 
 
Therefore, up to near the critical value, ${\cal R}_\phi$
and ${\cal R}_h$ are frozen to their values at horizon crossing, as
expected for nearly massless and decoupled fields, while ${\cal R}_N$
in Eq.~(\ref{RNinf}) {\it decreases} as the ratio $\dot N/N$
evolves in time. In order to estimate its value by the time inflation
ends, we use the fact that the ratio $Q_N/N$ remains approximately
constant, and given by Eq. (\ref{QN}) all along inflation. Therefore, 
the contribution ${\cal R}_N$ at the critical
value is given by Eq. (\ref{RNinf}) but evaluated using $(\dot N/N)_c$
in Eq. (\ref{dotNNc}),  
\be
{\cal R}_{Nc} \simeq 2 \left(\frac{N}{\dot N} \right)_c\left(\frac{\dot
h}{h} \right)_* {\cal R}_{h*} 
\simeq 2 \left(\frac{\eta_h}{\eta_\phi^{1/3}}
\right) \left(\frac{H^2}{\kappa^2 (\phi_c^+-\phi_c^-)
\phi_c^+}\right)^{1/3} {\cal R}_{h*} \,.
\label{RNc}
\ee
Thus, by the end of inflation the value of ${\cal R}_N$ is suppresed 
with respect to its value at the time of horizon crossing by a factor
\be
{\cal R}_{Nc} \approx \left(\frac{\eta_\phi H}{\kappa
\phi_c^+}\right)^{2/3} {\cal R}_{N*} \approx \left(\frac{\eta_\phi
\phi_c}{m_P}\right)^{2/3} {\cal R}_{N*} \,, 
\ee
which for $\eta_\phi \simeq 0.01$ and $\kappa \phi_c \simeq 1$ TeV,
with $H\simeq 10$ MeV ($\phi_c \simeq 10^{13} GeV$) would give a
suppresion factor approximately of the order of $10^{-5}$. 

\subsection{During the phase transition.}
After crossing the critical value, the ratio $Q_N/N$ still remains
practically constant until $N$ reaches the straight line trajectory
Eq. (\ref{straight}). Both evolutions, that of $N$ and its 
fluctuation, are now governed by the tachyonic instability in their
effective mass. This means that ${\cal R}_N \propto Q_N/\dot
N$ diminishes again by a factor $\dot N/N$, before the fields reach
the global minimum.  Let us denote the 
point when they reach the trajectory Eq. (\ref{straight}) by a
subindex ``+''.  Thus, the value of ${\cal R}_{N+}$ is  given by
\be
{\cal R}_{N+}= \left(\frac{ (\dot N/N)_-}{ (\dot N/N)_+}
\right)  {\cal R}_{Nc} 
= \left( \frac{ (\phi_c^+-\phi)_+}{ (\phi-\phi_c^+)_-} \right) {\cal R}_{Nc}  
\label{RN+}
\,,
\ee  
with $R_{Nc}$ given by Eq. (\ref{RNc}) above, and $(\phi-\phi_c^+)_-$ by
Eq. (\ref{deltaphic}). By using Eq. (\ref{Nexp}) and the condition
Eq. (\ref{straight}), we have $(\phi_c^+-\phi)_+ \approx
(\phi-\phi_c^+)_-$. Therefore, during the waterfall the value of
${\cal R}_{N}$ would only decrease at most by an order of magnitude. 
Numerically we found $(\phi_c^+-\phi)_+ \simeq 0.2 (\phi-\phi_c^+)_-$

In the final part of the transition from $\phi_c$ to $\phi_0$, 
the evolution of the quantum fluctuations become 
coupled through the interaction in the
potential. Given that $N=\sqrt{2} (\phi- \phi_c^+)$, and $\dot
N=-\sqrt{2} \dot \phi$,  for the fluctuations we have $Q_N
\propto \dot N$ and $Q_\phi \propto \dot \phi$ such that 
$Q_N= -\sqrt{2} Q_\phi$ \cite{us}; i.e, once
Eq. (\ref{straight}) is fulfilled, ${\cal R}_N$ and ${\cal R}_\phi$
remain constant. 
By the time they reach the global minimum we simply have:
\be
{\cal R}_{\phi0}={\cal R}_{N0}= {\cal R}_{N+} \,.
\ee
Because the background Higgs is still moving towards the minimum
but with $\dot h \ll \dot \phi$, the total curvature perturbation is 
then given by the contributions of the 2 singlets, with
\be
{\cal R}_0 \simeq {\cal R}_{\phi0}={\cal R}_{N0} \,.
\ee
Using Eqs. (\ref{RN+}) and (\ref{RNc}), we then have:
\be
{\cal R}_0 \simeq 0.4 \left(\frac{\eta_h}{\eta_\phi^{1/3}}\right)
\left(\frac{H^2}{\kappa^2 (\phi_c^+-\phi_c^-)\phi_c^+} \right)^{1/3}
{\cal R}_{h*}\,.
\label{R0}
\ee
And for the spectrum, $P_{{\cal R}} \equiv k^3 \langle | {\cal R} |^2
\rangle/(2 \pi^2)$, we obtain
\be
P^{1/2}_{{\cal R}_0} \simeq 0.4 
\left(\frac{H_*^2}{\kappa^2 (\phi_c^+-\phi_c^-) \phi_c^+} \right)^{1/3} 
\frac{H_*}{\eta_\phi^{1/3} h_*} \approx
\left(\frac{\phi_c}{m_P}\right)^{2/3} \frac{H_*}{\eta_\phi^{1/3} h_*}
\label{spectrum0}\,. 
\ee
This can be compared to the initial value of the amplitude of the
spectrum, given by that of the inflaton at horizon crossing,  
\be
P^{1/2}_{{\cal R}_*} \approx \frac{H_*}{\eta_\phi \phi_*} \,,
\label{standard}
\ee
with 
\be
P^{1/2}_{{\cal R}_0} \approx \left(\frac{\eta_\phi
\phi_c}{m_P}\right)^{2/3} \frac{\phi_*}{\ h_*} P^{1/2}_{{\cal R}_*} \,.
\ee
Clearly, Eq. (\ref{spectrum0}) can be larger than the standard
value Eq. (\ref{standard}), if $h_*/\phi_* \ll (\phi_c
\eta_\phi/m_P)^{2/3}$, which is 
required in any case to ensure slow-roll inflation.  
In this scenario, the total curvature at the end of inflation grows
from its value at horizon crossing, 
as the fields dynamic changes and the
relative contribution of the $N$ field to the total curvature become
non-negligible during the phase transition. 
As we have already mentioned, this can be interpreted as converting
the isocurvature perturbation between $N$ and $\phi$ into the
adiabatic one ${\cal R}$, see Eq. (\ref{dotRscalars}). However,
the presence during inflation of a large amplitude  
${\cal R}_N$ is  initially controlled by the Higgs.  Because of that we  
prefer to refer to the Higgs $h$ field as ``curvaton'', instead of the
mediating field $N$. 

The Higgs does not contribute per se to the total
curvature at this point, because its
velocity is still negligible with respect to the others. Nevertheless, the
evolution of $h$, $\dot h$ and $Q_h$ resembles that of the other
fields.  First, during the waterfall  when $N$ becomes larger than
$H^2/\phi$, the slow-roll approximation for the Higgs 
breaks down.  The evolution of the Higgs is then dominated by its
linear coupling to $N$, such that
\be
\ddot h \approx 2 \lambda h N (\kappa \phi -
\frac{A_\lambda}{\sqrt{2}}) \,,
\ee
and then $\dot h \propto\sqrt{N}$. The fluctuation $Q_h$ follows
the background field, and as a result $R_h$ decreases. Until the other
fields reach the the trajectory Eq. (\ref{straight}); then again we
recover the relation $Q_h
\propto \dot h$, and ${\cal R}_h$ will tend to 
level with the others.

\begin{figure}[t]
\begin{tabular}{cc} 
\epsfxsize=8cm
\epsfysize=8cm
\epsfbox{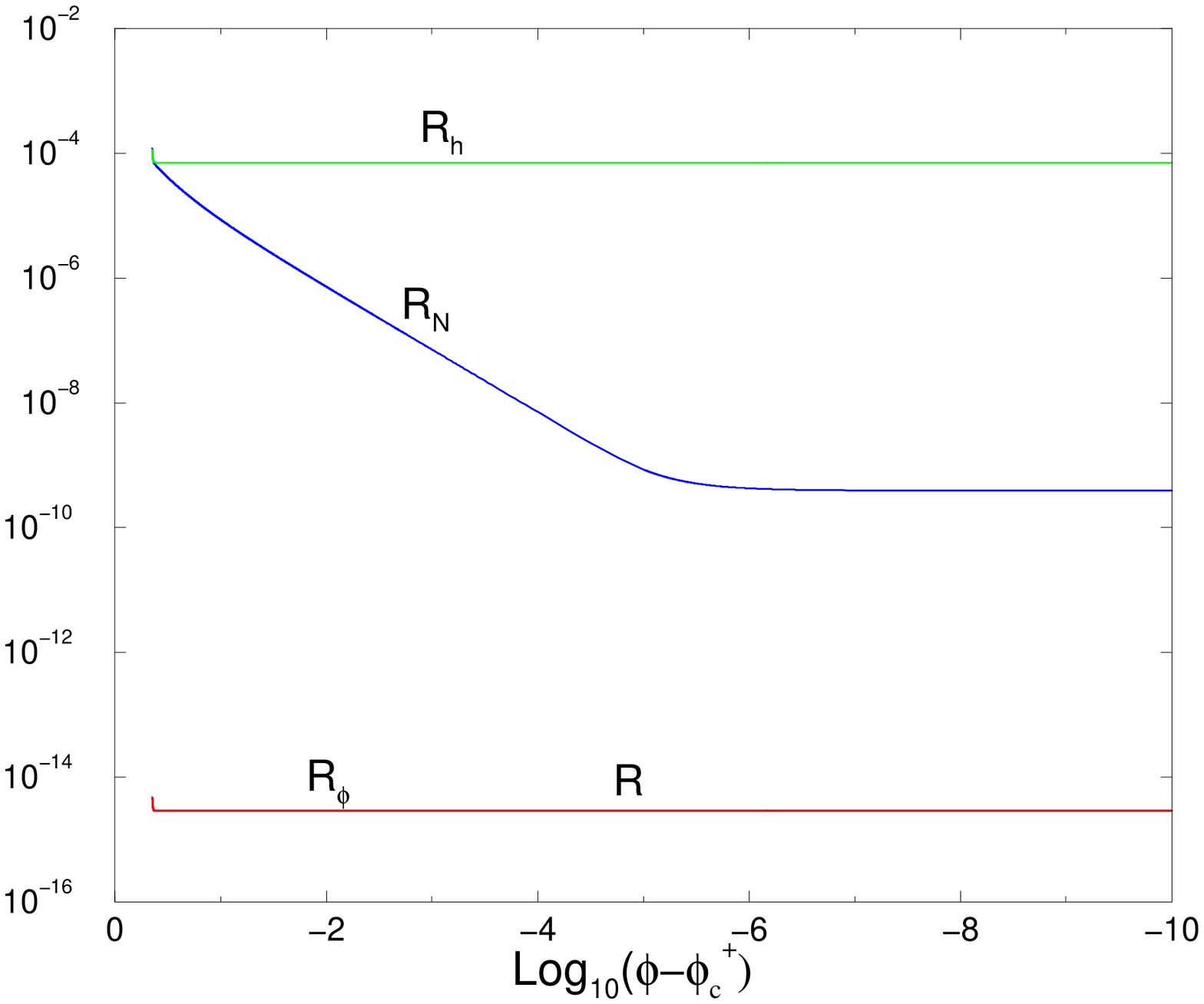}
&
\epsfxsize=8cm
\epsfysize=8cm
\epsfbox{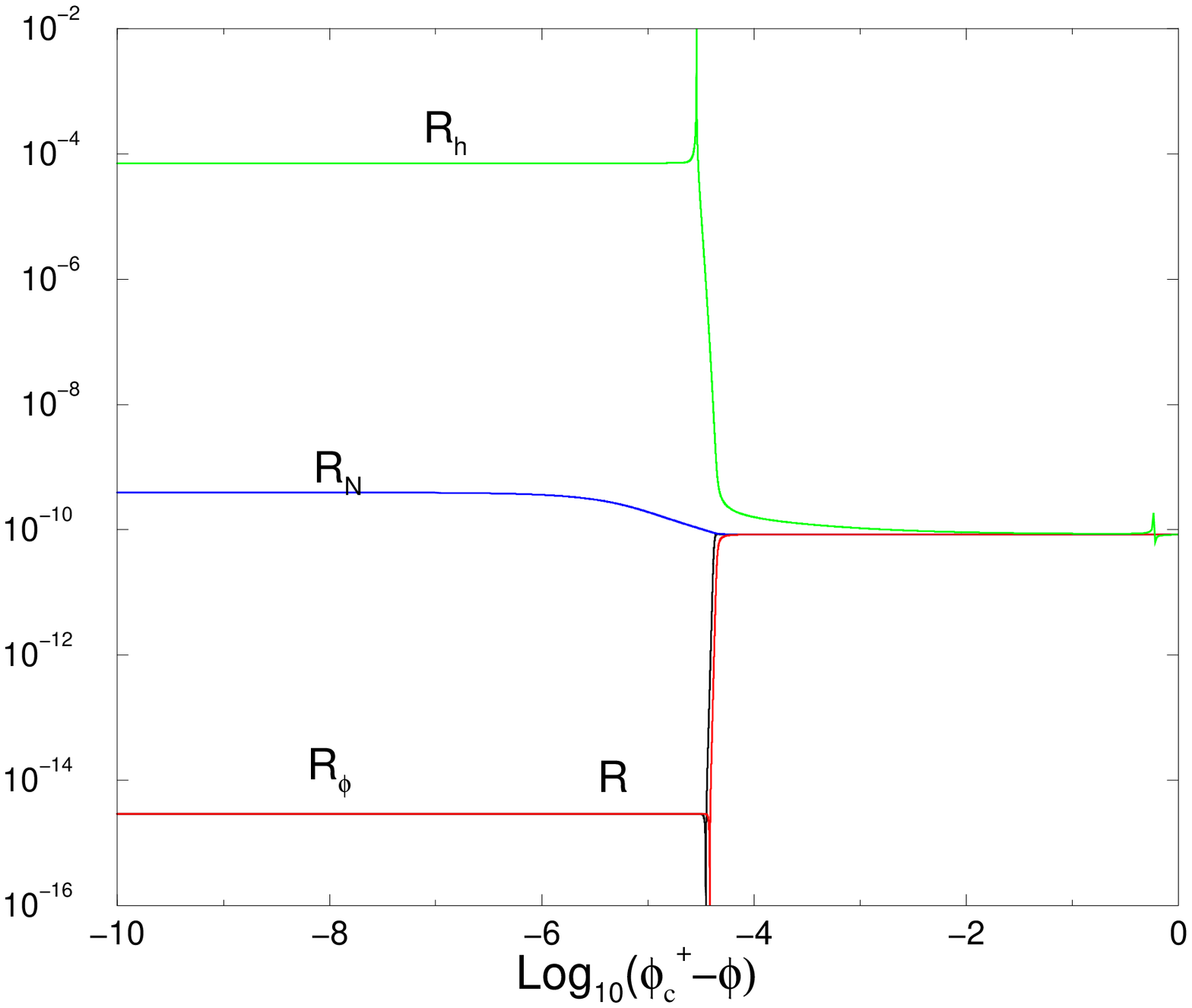} 
\end{tabular} 
\caption{\footnotesize Evolution of the spectrum $P^{1/2}_{{\cal
R}_\alpha}$ of the different 
curvature perturbations for each field, versus $\phi-\phi_c^+$, for
$k=aH$ 60 e-folds before the end of inflation. 
On the left plot it is shown their behaviour during inflation, and on the
right one from the critical value up to
the global minimum. Values of the parameters same than in
Fig.(\ref{fig0}). 
\label{fig1}}
\end{figure}

The behavior of the spectrum of the different contributions $R_\alpha$ during
inflation and the phase transition, is plotted in
Fig. (\ref{fig1}). The evolution is shown in logarithmic scale versus
the value of $\phi-\phi_c^+$, same than in Fig. (\ref{fig0}). 
In the LHS plot, 
we follow them from horizon crossing $k=aH$, taken to be 60 e-folds
before the end, through inflation, up to
the critical value (strictly speaking, at infinity outside  the
plot); and in the RHS plot we have their evolution during the
transition towards the global minimum.  This plot summarises the
previous findings:  during inflation ${\cal R}_h$ and ${\cal R}_\phi$ remains
constant, while the initial ${\cal R}_{N} \simeq {\cal R}_{h*}$ is
suppressed as it approaches the critical point due to the  increasing
$\dot N$ (see Fig. (\ref{fig0})). When inflation ends, $N$ falls fast
into the global minimum, dragging in the way also the inflaton and to
some extend the Higgs (Fig. (\ref{fig0})). As a result, their quantum
fluctuations will tend to follow each other, such that finally ${\cal
R}_{h0}\simeq {\cal R}_{N0}= {\cal R}_{\phi0}$.  

Nevertheless, given the model parameter values in section 2.1, it
seems that the 
amplitude of the spectrum in Eq. (\ref{spectrum0}) is too small to
match the COBE value, $P_{\cal R}^{1/2} \simeq 5 \times 10^{-5}$,
unless we fine-tune the value of $\eta_\phi$.  
However, this is the value at the end of inflation. This
will again be modified by the subsequent evolution of the fields
during reheating, in particular it can be amplified by ``preheating''
effects. 

\section{Curvature perturbation during reheating and preheating.} 

When the background fields oscillate around their
global minimum values, all the three
effective masses $\bar{m}_i$ become of the same order of magnitude, i.e.,
$\bar{m}_i\sim O(\kappa \phi_0) \sim O(1$ TeV),
this being the typical value for the frequency of the three
oscillating fields. Given the mixing in the potential, they will
end oscillating with similar amplitudes. That is, the vacuum
energy term $V(0)$ ends being equally redistributed among the three
matter fields. Later on, they will decay and transfer
that energy into radiation. This is the essence of the reheating
process. In our scenario, the Higgs
has the largest perturbative decay rate  $\Gamma_h$ due to either  its
large top Yukawa coupling, or gauge couplings $\alpha_W$, i.e.,
$\Gamma_h \simeq \alpha_W (\kappa \phi_0)$. On the other hand, the
singlet fields are 
very long-lived, due to their tiny Yukawa couplings, $\kappa \sim
\lambda \sim 10^{-10}$, with $\Gamma_{N,\phi} \simeq \kappa^2 (\kappa
\phi_0)$. Therefore, at least perturbatively, we expect the Higgs to be 
the first in being converted into radiation almost immediately after
the oscillations start. 
%Notice however that typically $\bar{m}_i
%>\Gamma_h \gg H$, so that indeed $h$ oscillates before decaying, but
%it does so in a tiny fraction of a Hubble time. No time for
%the expansion to redshift the energy densities, so practically the
%Universe enters the reheating era  with part of the vacuum energy 
%already converted into radiation (Higgs decay products). 
At the same time, Higgs perturbations are converted into those of the
radiation fluid.  

Nevertheless, previous to any perturbative decay, the evolution of the
system might be dominated by non-perturbative effects as those of
``preheating'', i.e., the parametric amplification of the fluctuations
in a background of oscillating fields \cite{preheating1,preheating2}. 
Through parametric resonance, induced
by the time dependent effective mass term in the evolution equations,
field mode amplitudes can grow exponentially in time within certain
resonance bands in $k$-space. The parametric resonance can be present
whenever there is a non-adiabatic change in the effective masses, 
$|\dot{\bar m}_i(t)/\bar m_i(t)| > 1$. 
Thus, preheating become a more effective
mechanism of transferring energy than the standard perturbative decay of
the fields, in this case from the oscillating background
fields to their quantum fluctuations. The question is whether it
also provides a different and efficient non-adiabatic source for 
the curvature perturbation, with 
super-Hubble ($k \ll aH$) fluctuations exponentially amplified during
preheating \cite{kodama2,basset,basset2,damping,counterexample}. This would 
translate into an exponential increase in the curvature perturbation. 
In hybrid inflation the effect
can be stronger first due to the presence of negative effective squared
masses at the end of inflation \cite{us,tachyonic}, which are per se a
source of instabilities in the evolution equations for the fluctuations.     
In addition the ``mixing'' through the potential  between the fields
can also enhance the resonance \cite{us,premixing}

\begin{figure}[t]
\epsfysize=10cm \epsfxsize=10cm \hfil \epsfbox{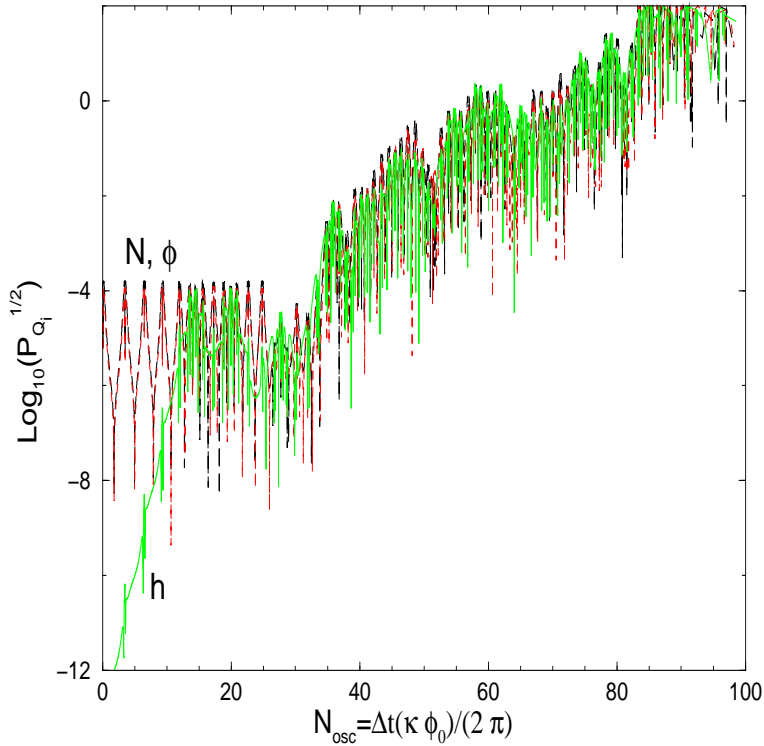} \hfil
\caption{\footnotesize Shown is the amplitude of the spectrum of the
field fluctuations versus $N_{osc}= \Delta t(\kappa \phi_0)/(2 \pi)$, 
without taking into account the decay of the Higgs fields.  
The numerical integration starts when the fluctuations left the Hubble
horizon, i.e., $k=H_* a_*$, roughly 60 
e-folds before the end of inflation. 
We have taken the values of the parameters as before:
$\phi_0=10^{13}$ GeV, $\phi_c^+=2 \phi_0$, $\kappa \phi_0=1$ TeV,
$A_\lambda=1.5 \sqrt{2} \kappa \phi_0$, $m_\phi=m_h=0.1 H$ and
$h_*=247$ GeV. 
\label{fig2}}
\end{figure}

Whether or not super-Hubble perturbations can be
parametrically amplified during preheating is a model-dependent
question 
\cite{damping,counterexample} that has to be studied case by case. 
We have then numerically integrated the evolution equations for fields
and fluctuations during the oscillatory period, in order to see the
effect on the curvature. As already mentioned, the presence of
tachyonic instabilities in the effective masses, combined with the
fact that the fields are oscillating with large amplitudes and the
effect of the expansion is negligible, could give rise to  a strong
parametric resonance effect just in a few oscillations. This can be
seen in Fig. (\ref{fig2}), where we have let the fields oscillate. It
can be noticed that the 
resonance for the super-Hubble fluctuations {\it only sets in once the Higgs
start oscillating} with an amplitude comparable to that of the others,
disturbing the straightline trajectory followed until then by $N$ and
$\phi$. Within that trajectory, their large scale field fluctuations with $k
\ll a H$ follow the same evolution equations as
their background fields, without amplification \cite{gordon2}.     

Although it may seem that the amplitude of the curvature perturbation
will increase beyond control, the above example was not realistic as
we still need to take into account the effect of the Higgs ``decay''. 
We remark that we used
the term ``decay'' in an ample sense, meaning the transfer or
dissipation of energy into radiation.
In general, parametric resonance occurs within certain $k$-bands, and
if not other, the backreaction of the quantum fluctuations will tend
to shut off the resonance. Being a non-perturbative effect, it is very difficult
to study the the whole process without resorting to numerical lattice
calculations \cite{tachyonic,lattice}, which are far beyond the scope of this
paper. Analytical or semi-analytical calculations in $k$ space have
been carried out using the Hartree-Fock or large N approximations
\cite{boyanosky}. Using those approximations, the overall effect of
the quantum fluctuations would be analogous to introducing
``friction'' in the evolution equations \cite{hosoya,berera}. Therefore, we
model this effect by using a constant {\it parameter} $\Gamma_h$,  
that is, 
a friction-like term in the Higgs evolution equation
\cite{kodama,preheating2},   
\be
\ddot h + 3 H \dot h + V_h = -\Gamma_h \dot h \,. 
\ee
$\Gamma_h$ gives then the typical time
interval needed for the conversion into radiation to take place.
The assumption behind is that the quantum fluctuations are first 
produced by the Higgs, this being the field with the larger coupling
to massless degrees of freedom. 
Although we are still working explicitly with the
background ``fields'' and their perturbation variables, radiation is
described as a perfect fluid component, with background energy density
$\rho_R$ and 
pressure $p_R=\rho_R/3$. From the conservation of the total (fields
plus radiation) stress-energy tensor, it follows 
\be
\dot \rho_R  + 3 H (\rho_R + P_R)= \Gamma_h \dot h^2= \Gamma_h (\rho_h+P_h)
\,,
\ee
where the RHS is the source term from the Higgs. 

Similarly for the Higgs fluctuations we have \cite{kodama}:
\be
\ddot Q_h + 3 H \dot Q_h + (\frac{k^2}{a^2}+ V_{hh}) Q_h + V_{h \phi}
Q_\phi + V_{h N} Q_N= -\Gamma_h \dot Q_h \,;
\label{ddotQh} 
\ee
here for simplicity we have neglected the sub-dominant metric
contributions. In addition, we need to introduce their perturbations
$\delta \rho_R$ 
and $\delta P_R$, in particular their corresponding gauge invariant
variables. Gauge invariant perturbations can be defined in 
different ways \cite{kodama}, and a summary with our choice 
of gauge-invariant variables and their evolution equations is given in
Appendix A.   
In order to compute the amplitude of the total curvature perturbation
${\cal R}$ 
we directly integrate the evolution equations for the scalar 
field fluctuations, and 
the contribution of the radiation to the curvature perturbation,
${\cal R}_R$ (see Eq. (\ref{Rr}) in Appendix A), such that 
\be
{\cal R}= \sum_\alpha \frac{\rho_\alpha +P_\alpha}{\rho +P} {\cal R}_\alpha
=\frac{1}{\rho+P}\left(\sum_{\alpha=\phi,N,h} H \dot \phi_\alpha Q_\alpha
 + (\rho_R + P_R) {\cal R}_R \right) 
\,. 
\label{curdefR}
\ee

\begin{figure}[t]
\begin{tabular}{cc} 
\epsfxsize=8cm
\epsfysize=8cm
\epsfbox{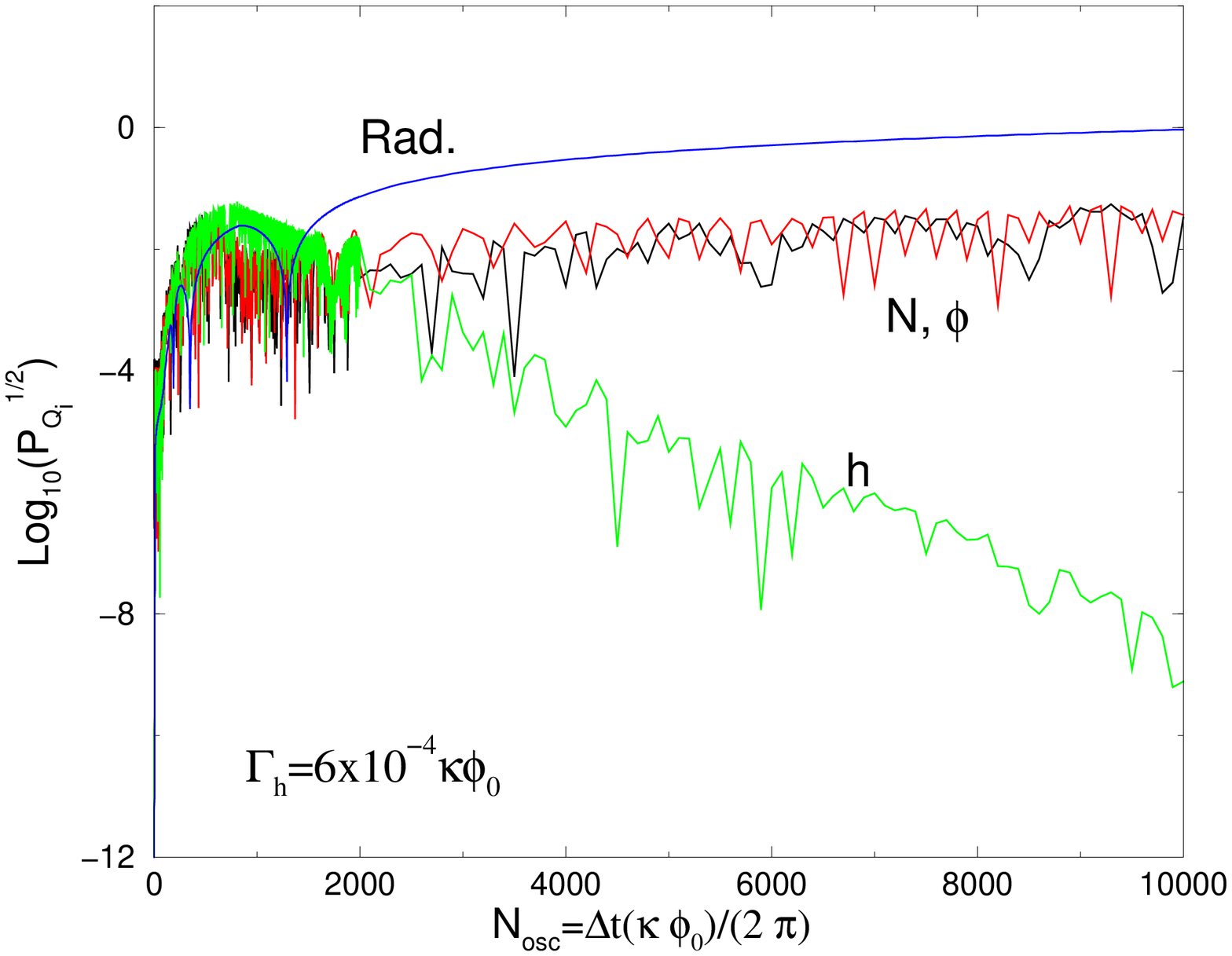}
&
\epsfxsize=8cm
\epsfysize=8cm
\epsfbox{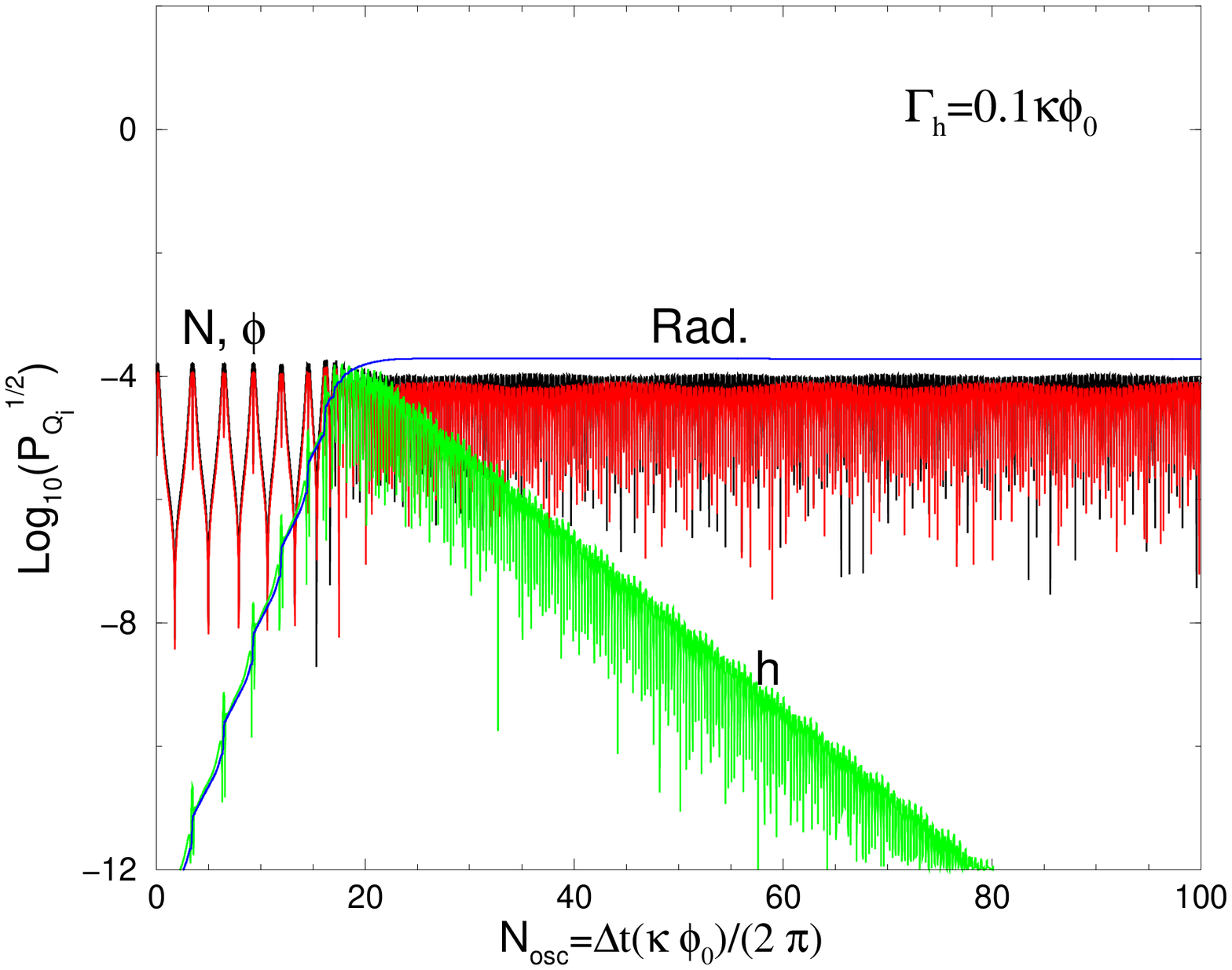} 
\end{tabular} 
\caption{\footnotesize Same than in Fig. (\ref{fig2}), but including
the decay of the Higgs, for two values of $\Gamma_h$. We also
include for comparison the analogous of the amplitude of the spectrum
for the radiation, define as $P_R = ((\rho_R+P_R)^{1/2}/H^2)
P_{{\cal R}_R}$. 
\label{fig3}}
\end{figure}    

By changing the patter of oscillations of the scalar fields, the
parameter $\Gamma_h$ will control not only what fraction of the total
energy density is transferred into radiation but how 
much ``preheating'' is allowed.  
As energy is transferred from the oscillating scalar fields to
radiation and the amplitude of the
oscillations decrease, this has the   
effect of shifting the resonance from the ``broad'' (large growth
index for the fluctuations) regime towards the
narrow (smaller growth) regime, until it finally dissapears. 
From Fig. (\ref{fig2}) we can read that preheating
built-in on a time interval of the order of $\Delta t_{preh} (\kappa
\phi_0) \simeq O(100)$. Then, depending on the
value of $\Gamma_h \Delta t_{preh}$ we divide the parameter range into
3 different regimes, ``large'', ``medium'' and ``small'', in order to
study the influence of the friction term in the final value of the
curvature:   

{\bf (a) Large: $\Gamma_h \Delta t_{preh} \gg 1$ ($\Gamma_h \approx O(\kappa
\phi_0) \approx O({\bar m}_i)$)}.   

The friction is too large to allow the Higgs to oscillate (the Higgs motion is
overdamped). The motion of the other two singlets is practically
undisturbed by the presence of the Higgs, and they oscillate in phase
along the straight line trajectory Eq. (\ref{straight}). Thus,  there
is no parametric resonance on the large scale perturbations. In
addition, only the energy density held by the Higgs during inflation
would be transfer into radiation, negligible compared to the total energy
density. Inflation ends with the vacuum energy being converted into
that of ``matter'' (oscillating fields), and there is no further
change in the curvature perturbation. The final value of the amplitude
of the primordial spectrum is then given by the value at the end of
inflation, Eq. (\ref{spectrum0}).
  
{\bf (b) Medium: $\Gamma_h \Delta t_{preh} \approx O(10 - 1)$}. 

The friction is large enough to quickly shift the resonance to the
narrow regime, but allowing first the 
Higgs to oscillate with an amplitude similar to the
others. In this range of values, the energy density in radiation is
comparable to the one left in the singlets system. 
As an example, on the RHS in Fig. (\ref{fig3})  
where we have plotted the amplitude for the spectrum
of the fields fluctuations for $\Gamma_h = 0.1 (\kappa \phi_0)$.   
The Higgs decays in a time interval
comparable to that of the frequency of the oscillations, and
comparable to the time needed to built the resonance for the field
fluctuations, so that indeed the 
latter hardly starts for the fields; Higgs fluctuations are just
converted into radiation perturbations. 

Therefore, very quickly we are left with a mixture of radiation and
two oscillating singlets. We may think, quite correctly, that in any
case given that the individual curvature perturbations were more or
less the same at the end of inflation, and the first stage of the
oscillations does not seem either to create large differences among
them, there is no further large entropy/isocurvature perturbation to be
converted into curvature one, so that the evolution of the later would
not be much altered. However, 
it is not only the presence of the radiation
which make the cosmic fluid non-adiabatic and may affect the evolution
of the curvature perturbation, but the fact that the fields are still
oscillating not in phase with large amplitudes. Taking now the system to
be made of 3 components, radiation, $N$ and $\phi$ field, the
evolution equation for the total curvature can be written as\footnote{
We neglect for simplicity terms that go like $(k/aH)^2 \ll 1$.}
\cite{kodama,silvia}:
\bea
\dot {\cal R} &\simeq& H (1-c_R^2) \frac{h_S h_R}{h_T^2} S_{SR} + 3 H (1
-c_S^2) \frac{h_S h_R}{h_T^2} ({\cal R}_S- {\cal R_R})\nonumber \\
&& -3 H (c^2_\phi - c^2_N) \frac{h_\phi h_N}{h_T h_S}
({\cal R}_\phi -   {\cal R}_N) 
\label{dotRcur}\,,
\eea
where we use the same notation than in Eq. (\ref{dotRscalars});
the subscript ``T'' means total quantities, and
``S'' refers to the combination of the two scalar fields, 
 with $h_S=h_\phi+h_N=\dot \phi^2+\dot N^2$ and  
\be
{\cal R}_S= \frac{h_\phi}{h_S} {\cal R}_\phi + \frac{h_N}{h_S} {\cal
R}_N \,;
\ee
$S_{SR}$ is the entropy
perturbation between ``scalars'' and ``radiation'', defined in 
Appendix A, Eq. (\ref{entropy}); and $c_\alpha^2=\dot p_\alpha/\dot
\rho_\alpha$ are the sound speed for each component. For the scalar
fields they are given in Eq. (\ref{calpha}),  
and they are time-dependent, fast oscillating functions. Therefore, even if on
average (over many oscillations) we could consider both fields as
behaving like matter, with $\langle c_\alpha^2 \rangle=0$, strictly 
speaking the last term in Eq. (\ref{dotRcur}) only cancels out when
$ V_N \dot \phi - V_\phi \dot N=0$, which is no more than the
condition for the straight-line trajectory. Pushed away from that
trajectory by the Higgs, the last term in Eq. (\ref{dotRcur})
introduces another {\it non-adiabatic source} in the equation, different
from the one in the first line due to the presence of the
radiation. Both together can still enhance the curvature perturbation
during the long reheating period that follows inflation. Until the
radiation is redshifted away, we are left again only with
``matter'', and the curvature remains constant thereon.   

{\bf (c) Small: $\Gamma_h \Delta t_{preh} < O(1)$.}
 
There is still partial preheating, while the system continuosly shifts
from broad to a narrow resonance, and the fluctuations are initially
amplified. In addition, small values of $\Gamma_h$ are more efficient
in ``extracting'' energy from the oscillating fields, so that by the
time the Higgs oscillations disappear we end practically with a radiation
dominated Universe, with $\rho_R/\rho_T \simeq 0.9$. 
As an example, on the LHS in Fig. (\ref{fig3})  
where we have plotted the amplitude for the spectrum
of the fields fluctuations\footnote{Numerically we found that for
$\Gamma_h=5 \times 10^{-4} (\kappa \phi_0)$, and 
the other value parameters as given in the figures, the
resonance remains broad for too long, given too large amplitude
fluctuations.} for $\Gamma_h = 6\times 10^{-4} (\kappa \phi_0)$. 
Still there is enough energy left for the scalar fields to oscillate,
so the last term in Eq. (\ref{dotRcur}) cannot be neglected.  

\begin{figure}[t]
\epsfysize=10cm \epsfxsize=10cm \hfil \epsfbox{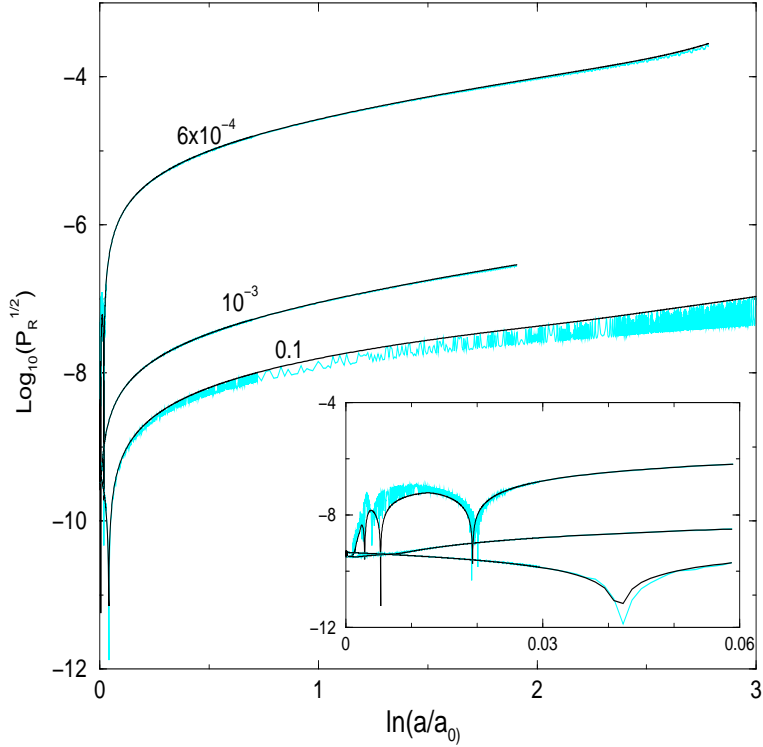} \hfil
\caption{\footnotesize Amplitude of the spectrum of the curvature
$P^{1/2}_{{\cal R}}$ (wiggly line)
and that of the radiation $P^{1/2}_{{\cal R}_R}$ (solid line), for
different values of $\Gamma_h$ given in units of $(\kappa \phi_0)$. 
Values of the other parameters as in Fig. (\ref{fig0}). 
\label{fig4}}
\end{figure}    

In Fig. (\ref{fig4}) it is shown the spectrum of the total curvature
perturbation and that of the radiation, for ``medium'' and ``small'' values of
$\Gamma_h$, but using now a more cosmological time scale $\ln(a/a_0)$,
with $a_0$ set when reheating/preheating (oscillations) starts. The
smallest $\Gamma_h$, the larger the initial amplification of the
fluctuations and then the curvature. After roughly half e-fold, the
evolution follows a more regular pattern, with the curvature
increasing as ${\cal R} \sim \ln(a/a_0)$, until the radiation is
diluted away and we recover ${\cal R} \simeq Constant$. 
%For example, for $\Gamma_h=0.2 (\kappa \phi_0)$, we have
%$(\rho_R/\rho_T)^{max}\simeq 0.45$, and then the radiation becomes
%subdominant by $\ln (a/a_0) \simeq 2$, and the curvature perturbation
%levels off as can be seen in the plot. 
For example, for $\Gamma_h=0.1 (\kappa \phi_0)$, we have
$(\rho_R/\rho_T)^{max}\simeq 0.78$, and then the radiation becomes
subdominant by $\ln (a/a_0) \simeq 3.5$, and the curvature perturbation
levels off as can be seen in the plot. 
The spread in the value of
$P_{\cal R}^{1/2}$ is related to the oscillations: 
during those the kinetic energy of the fields passes close to cero,
at which points the curvature is bounded by the contribution in
radiation. 

To summarise, we enter the reheating phase after inflation with a
non-adiabatic mixture of oscillating fields and radiation, which will
influence the evolution of the curvature perturbation (see
Eq. (\ref{dotRcur})). On average, 
it will increase among one and two orders of magnitude before the
radiation is redshifted away and becomes constant. In addition, for
small values of 
$\Gamma_h$ there is further enhancement of the perturbations due to
initial preheating effects, which are difficult to quantify
analytically. The final spectrum would be given by that at the end of
inflation, Eq. ({\ref{spectrum0}), but multiply by a  function of
$\Gamma_h$,  
\be
{\cal R} \sim F[\Gamma_h]_{reh} \left(\frac{\phi_c}{m_P}\right)^{2/3}
\frac{H_*}{\eta_\phi^{1/3} h_*} \,,
\ee 
which depending on the value of $\Gamma_h$ can be as large as
$10^{5}$, and it could then compensate for the suppression of the
initial spectrum of $N$ and $h$ during the phase transition at the end
of inflation.  

Before ending this section, some remarks about some of the assumptions
involved in the final value of the curvature perturbation. 
First, as already remarked in section 2.3, the value of the Higgs vev
is such that it never interferes with the evolution of the other fields
during inflation, see Eq. (\ref{condh}). Its {\it evolution} plays an
important r\^ole during the preheating stage, but not before. Were
this not the case, and the value of the Higgs background field $h$ is
such that the combination $\lambda h$ became larger than $H$
before the critical point this would signal the end of inflation, at
which point we would have already ${\cal R} \simeq {\cal R}_N \simeq
{\cal R}_h \simeq {\cal R}_\phi$, with a different (smaller)
suppression factor than that given in Eq. (\ref{R0}). In this case $h$
would be the field with the increasing kinetic energy, so that it
would be directly the isocurvature perturbation between the inflaton and $h$
which sources the total curvature perturbation. Nevertheless, because
in this case the initial value of the Higgs vev would be larger, its
initial curvature value at horizon crossing becomes smaller, and we do
not obtain a larger amplitude for the curvature at the end of the
phase transition. 

For the range of values $h_*$ for which Eq. (\ref{spectrum0}) holds, 
the final value of the spectrum will scale as the inverse of the
Higgs' vev, or almost. For ``medium'' values of $\Gamma_h$ this is the
case, as reducing or increasing the value of $h_*$ has null or little
effect on the patter of oscillations. However, for ``small'' values of
$\Gamma_h$ for which preheating is at hand, any change in the
parameters may affect the final value, by shifting the resonance from
one band to another, from broad to narrow. This will modify the
initial ``boost'' in the perturbations, but not the afterwards
behaviour ${\cal R} \propto \ln(a/a_0)$.   

\section{Discussion}

Like in the original curvaton model, the final primordial amplitude of
the spectrum is controlled by  the Higgs vev $h_*$ instead of
the inflaton vev at the time of horizon crossing. This is achieved through
the $N$ field in hybrid inflation, which  mediates between the
inflaton and the Higgs/curvaton flat direction. In addition, due to 
dynamics of the system, the curvature perturbation changes first
during the phase transition, which enhances the initial amplitude at
horizon crossing by a factor  $\sim (\eta_\phi
\phi_c/m_P)^{2/3}(\phi_*/h_*)$; and then
through preheating effects, which depend on the parameter $\Gamma_h$
but nevetheless give rise to an enhancement of the spectrum. For 
small values of $\Gamma_h$, both effect
can cancel each other such that one recover a similar curvaton
scenario result, with $P_{\cal R}^{1/2} \approx H_*/\eta_\phi^{1/3} h*$.
However, even in this simplify estimation, the different origin of the 
spectrum compared to the more standard curvaton scenario can be seen
in the factor $\eta_\phi^{1/3}$. That is, the amplitude of the
spectrum depends to some extend on the inflationary dynamics, in
particular on the mass of the inflaton field instead of the mass of
the Higgs field. As far as the Higgs evolution does not backreact onto
the others until the fields start oscillating (see Eq. (\ref{condh})),
it is only its vev at horizon crossing which sets the scale for the
amplitude.   
  
On the other hand, the evolution of the Higgs field and not the others
will control the spectral index $n_S$ of the primordial spectrum, 
\be
n_S-1 = \frac{ d \ln P_{\cal R}}{d \ln k} \,.
\ee
Although in the final value of the spectrum we have to take into
account the function $F[\Gamma_h]_{reh}$ which numerically may depend
on the value of $h*$, this does not bring any additional $k$
dependence into the spectrum. That function parametrise the effects on
the large scale spectrum during the oscillatory phase of the fields,
which depends largely on the patter of the oscillations and modify
equally all the wavelenths with $k \ll aH$ by the time of reheating.  
Given than in this model the variation of the Hubble parameter during
inflation is practically negligible, with $ 2 \dot H/H^2 \simeq
\eta_\phi^2 (\phi_c/m_P)^2$, and using $d \ln k \simeq H dt$, we simply have: 
\be
n_S-1 \simeq 2 \eta_h \simeq \frac{2 m_h^2}{3 H^2} \,,
\ee    
and practically no running of the spectral index. The recent WMAP data
\cite{wmap}, combined with other CMB experiments, 
prefers a red tilted spectrum with $n_S \simeq 0.96 \pm
0.02$. This leads to the constraint $|\eta_h| \leq 0.03$, which would only
require $m_h$  slightly smaller than $H$, $m_h \sim 0.3 H$. Strictly
speaking, the spectrum is red tilted when $\eta_h <0$, i.e., $m^2_h
<0$ during inflation. Although up to now we have been implicitely
assuming the squared Higgs mass parameter in the potential to be
positive, none of the results regarding the amplitude of the spectrum
would change had we taken initially the opposite sign.

\section{Summary} 

We have presented a variant of the curvaton scenario in SUSY hybrid
inflation where: 
(a) it is coupled to the inflaton through the mediating field in
hybrid inflation;  
(b) instead of being a late-decaying scalar, the curvaton decays
before the inflaton, already  at the beginning of the reheating
period.   
The essence
of the scenario is the same: isocurvature (relative) perturbations
between the fields originated during inflation can be converted into
adiabatic ones later on. Thus the primordial spectrum of perturbations
is given in terms of the parameters related to the curvaton field instead of
those of the inflaton. The conversion occurs when the curvaton energy
density becomes dominant or comparable to that of the inflaton. 

However, in our scenario it is  the fact that the fields are
coupled which first induced the conversion of the isocurvature
perturbation between the inflaton and the $N$ field into the total
one, previous to any decay. First, the coupling to the ``curvaton''
gives rise to an $N$ curvature perturbation given at the time of
horizon crossing by the value of the curvaton one. During the phase
transition from the false 
vacuum towards the global minimum, this contribution comes to
dominate the total curvature when the kinetic energy of the $N$ field
grows until becoming comparable to that of the inflaton (see
Fig. (\ref{fig1})). Therefore, the initial amplitude of the total
curvature perturbation at the time of horizon crossing, $P_{{\cal
R}_*}^{1/2} \simeq H_*/\eta_\phi \phi_*$, gets enhanced by a factor of
the order of $(\eta_\phi \phi_c/m_P)^{2/3} (\phi_*/h_*)$, already at
the end of inflation. 

Second, the curvaton coupling to the other fields is relevant during
the period of oscillations and it gives rise to preheating or
parametric amplification of the large scale fluctuations. In standard
SUSY hybrid inflation, the background fields oscillate along a  practically
straight-line trajectory. This regular behaviour prevent any
amplification of the fluctuations with wavenumbers $k \ll
Ha$. Including the coupled curvaton directly perturb the background
trajectory, which immediate 
effect is the amplification of the large scale fluctuations, and
consequently the amplitude of the curvature perturbation. Without a
detail numerical computation including backreaction and rescattering
effects of the modes (large and small scale) produced, it is difficult
to estimate how large could be this effect. However, heuristically we
expect the curvaton to quickly decay in a few oscillations, and this would
ensure that the resonance ends by the time the curvaton dissapears, if
not before. We model then this behaviour by introducing a ``decay rate''
term in the equation of motion of both curvaton background field and
fluctuations, which simply parametrises the energy transfer rate
between the curvaton and radiation. In a sense this is equivalent to
take into account the backreaction effect of the
fluctuations. Friction or dissipation effects are ultimately due to
quantum fluctuations, which modify the propagation of the fields. The
approximation behind our equations is that the dominant effects are due to
the fluctuations generated by the curvaton, with the largest decay
rate, but not directly coupled to the singlets, and then  we do not
consider any other direct backreaction effect in the evolution equations for
the singlets. Within this approximation, the smaller the decay rate
parameter is, the larger the initial  preheating and the larger the
final value of the amplitude of the curvature. Therefore, depending on
the value of $\Gamma_h$, the amplitude enhancement during
preheating can be larger than during the phase transition. 

This general behaviour of the curvature perturbation does not depend
on model parameters, but on the fact that we have coupled
fields. However, the model introduced in Section 2 was motivated by
particle physics considerations, which lead us to identify the
``curvaton'' D-flat direction with the Higgs of the MSSM. The $N$
field then originates the $\mu$ term in the superpotential, needed for
electroweak symmetry breaking. We also impose an $U(1)$ Peccei-Quinn
symmetry wich solves the strong CP problem. Those physical
requirements fixed uniquely the values of the parameters in the model,
as given in Section 2. The numerical results in the Figures have been
obtained using those values, mainly $\kappa \phi_0 \simeq \kappa
\phi_c \simeq 1$ TeV and $\phi_0 \simeq \phi_c \simeq 10^{13}$ GeV. Although
the vev of the Higgs D-flat direction during inflation can be
considered a free parameter, it looks natural to assume a value not
that far from its value at the electroweak symmetry breaking minimum,
i.e., of the order of  1 TeV. For that choice, it is clear that the
value of the spectrum at the end of inflation, Eq. (\ref{R0}),
although larger than at the time of horizon crossing, it is still far
below the observational bound $P_{\cal R}^{1/2} \simeq 5\times
10^{-5}$. However, as we have seen, preheating
effects can further 
enhanced this value. Although numerically the final value will depend
on the particular value of the parameter $\Gamma_h$, i.e., on how
fast  energy is transferred to the radiation, it can be seen in
Fig. (\ref{fig4}) that for example for $\Gamma_h \sim 1 $ GeV the
amplitude is of the correct order of magnitude.     

In conclusion we have discussed the potentially important r\^{o}le played
by preheating in certain variants of the curvaton mechanism
in which isocurvature perturbations of a D-flat (and F-flat) direction
become converted to curvature perturbations during reheating.
We have analysed the
transition from inflation to reheating in some detail, including the
dynamics of the coupled curvaton and inflation fields during this
transition. We have discovered that preheating could be an important
source of adiabaticity 
where parametric resonance of the isocurvature components amplifies
the super-horizon fluctuations by a significant amount.
As an example of these effects we developed 
a particle physics motivated model which we recently introduced in
which the D-flat direction is identified with the usual Higgs field. Our new
results show that it is possible to achieve the correct curvature
perturbations for initial values of the curvaton fields of order the
weak scale. 
In this model we have showed 
that the prediction for the spectral index of the final
curvature perturbation only depends on 
the mass of the curvaton during inflation,
where consistency with current
observational data requires the ratio of this mass to the
Hubble constant to be $\leq 0.3$.

\section*{Acknowledgments}

The authors are grateful to David Lyth for useful comments. MBG thanks
N. Antunes and J. Urrestilla for their useful comments and help with
the numerical calculations.  VDC was
supported at the University of Southampton by PPARC rolling grant and
at the University of  Oxford by PPARC rolling grant under the grant
number PPA/G/0/2002/00479.  

\section*{Appendix A} 

In this Appendix we summarise our conventions and notation for the
perturbations  and their evolution equations. We work with gauge
invariant (GI) quantities, to first order in linear perturbation
theory.   

$\bullet$ {\bf Definitions}: Linear scalar
perturbations of the metric are given by the line element: 
\be
ds^2= -(1+2 A) dt^2 + 2 a  B_{i} dx^i dt  \nonumber \\
      + a^2 [ (1-2 \psi) \delta_{ij} + 2 E_{,ij}] dx^i
dx^j \,, 
\ee
where $B_i= i k_i/k B$, and $E_{,ij}= (- k_i k_j/k^2) E$ for
scalar perturbations; $\psi$ is the gauge-dependent curvature
pertubation. On the other hand, the GI variable 
\be
\Phi =  -\psi + Ha (B- a \dot E) \,,
\ee
correspond to the curvature  perturbation in the longitudinal (or
zero-shear) gauge. 

The perturbations of the stress-energy tensor are: 
\bea
\delta T^0_0 &=& -\delta \rho \,,\\ 
\delta T^0_j &=& (\rho+P)(v_j- B_j) \,, \\
\delta T^j_0 &=& -(\rho+P)v^{j} \,,\\
\delta T^i_j &=& \delta P \delta^i_j + P \Pi^i_j \,, 
\eea
where $\Pi_{ij}$ is the anisotropic stress taken to be zero hereon. 
The comoving curvature perturbation is then given by: 
\be
{\cal R} = \psi + \frac{Ha}{k} (v - B)= -\Phi + \frac{Ha}{k} V \,, 
\ee
where $V= v - a k \dot E$ is the GI velocity of the fluid. 

Quantum scalar fields are decomposed as $\Phi_\alpha(x,t)=
\phi_\alpha(t) + \delta \phi_\alpha(x,t)$, where $\delta
\phi_\alpha(x,t)$ are the  quantum fluctuations of the field 
around background values $\phi_\alpha$. For a
multi-scalar field system, we have then:
\be
(\rho+P) ( v - B) = \frac{k}{a} \sum_\beta \dot \phi_\beta \delta
\phi_\beta \,,
\ee
and the comoving curvature perturbation:
\be
{\cal R} = \psi + \frac{H}{\rho +P} \sum_\beta \dot \phi_\beta
\delta \phi_\beta = \frac{H}{\rho+P} \sum_\beta (\rho_\beta+P_\beta)
Q_\beta \,,
\ee
where $Q_\beta$ are the Sasaki-Mukhanov GI variables:
\be
Q_\alpha= \delta \phi_\alpha + \frac{\dot \phi_\alpha}{H} \psi
\ee

$\bullet$ {\bf Evolution equations}: Let us consider a  multi-scalar system,
with the fields interacting through the potential
$V(\phi_\alpha)$. The derivaties of the potential with respect to the
fields are denoted by $V_\alpha= \partial V/\partial \phi_\alpha$, and 
$V_{\alpha \beta}= \partial^2 V/\partial \phi_\alpha \partial
\phi_\beta$. We also allowed  source terms $S_\alpha$ in
the evolution equation for the background fields, 
\be
\ddot \phi_\alpha + 3 H \dot \phi_\alpha + V_\alpha= -S_\alpha \,,
\ee
subject to the constraint $\sum_\alpha S_\alpha \dot \phi_\alpha=0$,
as required by the conservation of the total stress-energy tensor. The
corresponding GI source perturbation is defined as:
\be
{\cal D}S_\alpha= \delta S_\alpha + \frac{\dot S_\alpha}{H} \psi\,,
\ee
and the evolution of the fields perturbations is given by \cite{kodama}: 
\bea
\ddot Q_\alpha + 3 H \dot Q_\alpha + \frac{k^2}{a^2}Q_\alpha + 
& & \!\!\!\!\!\!\!\!\!\sum_\beta
V_{\alpha \beta} Q_\beta \\ \nonumber
&=& -{\cal D}S_\alpha - \left(\frac{k}{a}\right)^2 \dot \phi_\alpha
\frac{\Phi}{H} + \dot \phi_\alpha \dot {\cal A} + 2 (\ddot
\phi_\alpha+ 3 H \dot \phi_\alpha) {\cal A} \,,
\eea
where
\be
{\cal A} = A+ \left[ \frac{\psi}{H} \right]^\cdot= \frac{3}{2}
\frac{ \rho + P}{\rho} {\cal R} \,.
\ee
In the absence of source terms, the above equation reduces to the more
familiar one \cite{taruya}:
\bea
\ddot Q_\alpha + & &\!\!\!\!\!\!\!\!\!
3 H \dot Q_\alpha + \frac{k^2}{a^2}Q_\alpha +
\sum_\beta
V_{\alpha \beta} Q_\beta= 
\left[ \frac{1}{a^3 m^2_P} \left( \frac{a^3}{H} \dot \phi_\alpha \dot
\phi_\beta \right)^{\cdot } \right] Q_\beta\\
&=& \frac{3H}{\rho} \sum_\beta \left[ 3 H \dot \phi_\alpha \dot
\phi_\beta +\ddot \phi_\alpha \dot \phi_\beta
+\dot \phi_\alpha \ddot \phi_\beta -\frac{\dot H}{H} \dot
\phi_\alpha \dot \phi_\beta \right] Q_\beta \,.
\eea

The other particular case we have considered in this paper is when 
only one of the fields $\gamma$ decay into radiation
$\rho_R$, such that
\bea
\ddot \phi_\gamma + 3 H \dot \phi_\gamma + V_\gamma &=& -\Gamma_\gamma
\dot \phi_\gamma \,,\\
\dot \rho_R + 3 H (\rho_R + P_R) &=& \Gamma_\gamma \dot
\phi_\gamma^2=\Gamma_\gamma (\rho_\gamma + P_\gamma) \,. 
\eea
The GI radiation perturbations associated to the  energy density,
$\rho_R \Delta_{gR}$, and comoving curvature, ${\cal R}_R$, are then:
\bea
\rho_R \Delta_{gR} &=& \delta \rho_R +\dot \rho_R \frac{\psi}{H} \,,\\
{\cal R}_R &=& -\Phi + \frac{Ha}{k}V_R \label{Rr}\,.
\eea
The variables $Q_\alpha$ and $\rho_\alpha \Delta_{g\alpha}$ are the
fluctuations of the field and energy density respectively on uniform
curvature slices of space-time. For a scalar field, we have:
\be
\rho_\alpha \Delta_{g\alpha}= - \dot \phi_\alpha^2 {\cal A} + \dot
\phi_\alpha \dot Q_\alpha + V_\alpha Q_\alpha \,.
\ee

The evolution equations for the fluctuations can be written as:
\bea
\ddot Q_\alpha + & & \!\!\!\!\!\!\!\!\!3 H \dot Q_\alpha +
\frac{k^2}{a^2}Q_\alpha +\sum_\beta 
V_{\alpha \beta} Q_\beta 
=-\Gamma_\alpha \dot Q_\alpha \nonumber \\
&&+ \frac{3 H}{2 \rho} \left \{\sum_\beta 
\left[(2 \ddot \phi_\alpha + 3 H \dot \phi_\alpha + \Gamma_\alpha \dot \phi_\alpha)
\dot \phi_\beta + (2 \ddot \phi_\beta + 3 H \dot \phi_\beta + \Gamma
\dot \phi_\beta) \dot \phi_\alpha - 2 \frac{\dot H}{H} \dot
\phi_\alpha \dot \phi_\beta \right] Q_\beta \right.\nonumber \\
& & \left.+ (2 \ddot \phi_\alpha + \Gamma_\alpha \dot \phi_\alpha) 
\frac{ h_R {\cal R}_R}{H} - 2 \frac {\dot H}{H} \dot \phi_\alpha
\frac{ h_R {\cal R}_R}{H} + \dot \phi_\alpha (c_R^2-1) \rho_R
\Delta_{gR} + \dot \phi_\alpha \Gamma_\gamma \dot \phi_\gamma Q_\gamma
\right\}\,,
\eea
where we use the shorthand notation $h_\alpha= \rho_\alpha +
P_\alpha$. The equations for the radiation variables are given by:
\bea
(h_R {\cal R}_R)^\cdot&=& -\frac{\dot H}{H} h_R ( {\cal R} -  {\cal
R}_R)  + H c_R^2 \rho_R \Delta_{gR} - 3 H h_R {\cal R}_R
+\Gamma_\gamma \dot \phi_\gamma^2 {\cal R}_\gamma \label{dotrr}\\
(\rho_R \Delta_{gR})^\cdot&=& -3 H (1 + c^2_R) \rho_R \Delta_{gR} -
\frac{k}{a} h_R V_R \nonumber \\
&& +\Gamma_\gamma \dot \phi_\gamma ( 2 \dot Q_\gamma +\frac{\dot H}{H^2}
\dot \phi_\gamma {\cal R}) \,.
\eea  

The entropy perturbation $S_{\alpha \beta}$ introduced in
Eq. (\ref{dotRcur}) is defined as the difference between the energy
density perturbations:
\be
S_{\alpha \beta} = \frac{\rho_\alpha \Delta_{g\alpha}}{h_\alpha}
-\frac{\rho_\beta \Delta_{g\beta}}{h_\beta} \,,
\label{entropy}
\ee
which for non-interacting components is equivalent to 
\be
S_{\alpha \beta} = -3 H \left( \frac{\delta \rho_\alpha}{\dot \rho_\alpha}-
\frac{\delta \rho_\beta}{\dot \rho_\beta} \right) \,.
\ee

\section*{Appendix B}

Following Ref. \cite{gordon2}, for a multi-scalar system the field
perturbations can be rotated in field space into the ``adiabatic''
fluctuations along the background trajectory, which originates the total
curvature perturbation, and the ``entropy'' components, which are
ortogonal to the fields trajectory. 
Suppose we have $N$ numbers of fields an let us denote by $\sigma$ the
adiabatic field and $\theta_i$ the $N-1$ angles that parametrize the
trajectory.  ${\boldmath v}_\sigma$ will be the tangent vector to the field
trajectory and ${\boldmath v}_{\theta_i}$ the $N-1$ ortogonal vectors. Denoting by
$\boldmath Q$ the $N$-component vector with the fluctuations of the
original fields, the adiabatic and the $N-1$ entropy fluctuations are given by
\be
Q_\sigma = {\boldmath v_\sigma} \cdot {\boldmath Q}\,,\;\;\;\;\;\;\;
Q_{\theta_i} = {\boldmath v_{\theta_i}} \cdot {\boldmath Q}\,,
\ee
with  
\be
{\cal R} = H \frac{Q_\sigma }{\dot \sigma} \,,    
\ee
where $\dot \sigma^2 = \sum_{\alpha=1,N} \dot \phi_\alpha^2$. 
The evolution of the curvature for scales $k \ll aH$ can be written as
\be
\dot {\cal R} \simeq \frac{2 H}{\dot \sigma} \sum_{i=1,N-1} \dot
\theta_i Q_{\theta_i} \,,
\label{dotRtheta}
\ee 
This way, it is clear that the
evolution of the the total curvature depends only on the entropy
fluctuation fields, and the curvature of the background trajectory in
field space. As remarked in \cite{gordon2}, if the fields move in a
straight line in field space, $\dot \theta_i =0$, the curvature
perturbation on super-Hubble scales remains constant\footnote{This strictly
holds for scalar fields with minimal kinetic terms. Non-canonical
kinetic terms typically appear in extended gravity theories, and can
be the origin of the entropy source in the evolution of the curvature
\cite{juandavid,deruelle,curvaton3}}. 

In our scenario, with 3 scalar fields, the change in the curvature
during the phase transition can be related to the change in the
trajectory in the $\phi - N$ subspace, such that the corresponding
$\dot \theta_i$ becomes non-negligible and dominant. As far as the
Higgs does not contribute, the inflaton and $N$ will end moving in a
(practically) straight line, the system effectively reduces to that of
one scalar field, and $\cal R\simeq $ Constant. Later on, when the Higgs
``curves'' the trajectory, the evolution of ${\cal R}$ is
modified again.

\end{document}